# Rheology of mixed solutions of sulfonated methyl esters and betaine in relation to the growth of giant micelles and shampoo applications

Veronika I. Yavrukova [a], Gergana M. Radulova [a], Krassimir D. Danov [a],
Peter A. Kralchevsky [a,*], Hui Xu [b], Yee Wei Ung [b], and Jordan T. Petkov [b,1]

[a] *Department of Chemical and Pharmaceutical Engineering, Faculty of Chemistry and Pharmacy, Sofia University, 1164 Sofia, Bulgaria*

[b] *KLK OLEO, KL-Kepong Oleomas Sdn Bhd, Menara KLK, Jalan PJU 7/6, Mutiara Damansara, 47810 Petaling Jaya, Selangor Dalur Ehsan, Malaysia*

───────────────

* Corresponding author. *E-mail address*: pk@lcpe.uni-sofia.bg (P.A. Kralchevsky).

[1] Present address: Arch UK Biocides Ltd., Hexagon Tower, Crumpsall Vale, Blackley, Manchester M9 8GQ, UK

**Abstract.** This is a review article on the rheological properties of mixed solutions of sulfonated methyl esters (SME) and cocamidopropyl betaine (CAPB), which are related to the synergistic growth of giant micelles. Effects of additives, such as fatty alcohols, cocamide monoethanolamine (CMEA) and salt, which are expected to boost the growth of wormlike micelles, are studied. We report and systematize the most significant observed effects with an emphasis on the interpretation at molecular level and understanding the rheological behavior of these systems. The experiments show that the mixing of SME and CAPB produces a significant rise of viscosity, which is greater than in the mixed solutions of sodium dodecyl sulfate and CAPB. The addition of fatty alcohols, CMEA and cationic polymer, leads to broadening of the synergistic peak in viscosity without any pronounced effect on its height. The addition of NaCl leads to a typical salt curve with high maximum, but in the presence of dodecanol this maximum is much lower. At lower salt concentrations, the fatty alcohol acts as a thickener, whereas at higher salt concentrations – as a thinning agent. Depending on the shape of the frequency dependences of the measured storage and loss moduli, $G'$ and $G''$, the investigated micellar solutions behave as systems of standard or nonstandard rheological behavior. The systems with standard behavior obey the Maxwell viscoelastic model (at least) up to the crossover point ($G' = G''$) and can be analyzed in terms of the Cates reptation-reaction model. The systems with nonstandard rheological behavior obey the Maxwell model only in a restricted domain below the crossover frequency; they can be analyzed in the framework of an augmented version of the Maxwell model. The methodology for data analysis and interpretation could be applied to any other viscoelastic micellar system.

*Keywords*: Sulfonated methyl esters; Shampoo formulations; Wormlike micelles; Rheology of micellar solutions; Reptation-reaction model.

**Contents**

## 1. Introduction

Formation of large micellar aggregates of different morphology is most frequently observed in mixed surfactant solutions, in which the micelles are multicomponent and polydisperse in size [1-7]. Upon variation of solution's composition, high peaks in viscosity have been often observed [8-12]. Such concentration dependencies are of primary importance for various practical applications, e.g. for personal and household care detergency (e.g. shampoos and liquid detergents), because they allow one to control the micelle growth and formulation's viscosity [13-15]. The highest viscosities are observed in the presence of giant wormlike surfactant micelles. The high viscosity is due to the interplay of various processes and interactions that take place in such concentrated and internally structured solutions. First, the high aspect ratio of the long micellar aggregates gives rise to purely *hydrodynamic* interactions [16,17]. Second, de Gennes [18] identified the main relaxation mechanism for long linear polymers with *reptation*, which is related to the curvilinear diffusion of linear unbreakable molecules confined by their neighbors. Furthermore, in the case of long micellar aggregates ("living" polymers) Cates and coauthors [19-25] developed statistical theory, which accounts for micelle *reversible scission* and *end-interchange* processes. This theory predicts correctly the variations of zero-shear viscosity and other rheological parameters at not too high surfactant and salt concentrations [26]. Recently, Hoffmann and Thurn [27] took into account the energy of *sticky contacts* between micelles, which include contributions from the van der Waals, electrostatic, hydrophobic and hard-core interactions. Depending on the chemical nature of the component, whose concentration is varied, one could distinguish three types of viscosity peaks:

First, the variation of the *mole fractions* of the two basic surfactants (anionic and cationic, or anionic and zwitterionic) leads to a maximum in viscosity because of synergistic interactions between the two surfactants that promote growth of large self-assembled aggregates, usually – wormlike micelles, to the left of the maximum and diminishing of their size to the right of the maximum [10,28-31]. At that, the maximal viscosity corresponds to the concentration domain with the longest entangled micelles. The synergism can be due to favorable headgroup interactions (e.g. in a catanionic pair) [32,33], as well as to a mismatch in the surfactant chainlengths [34-36].

Second, in systems containing ionic surfactants the dependence of viscosity on the concentration of *added salt* (the so called salt curve) often exhibits a high peak [28,37-43]. In this case, the peak could be explained with a transition from wormlike micelles to branched micelles [44-48]. The initial growth of wormlike micelles could be explained with the screening of the electrostatic repulsion between the surfactant headgroups by the electrolyte, whereas the subsequent transition to branched aggregates can be interpreted in terms of surfactant packing parameters and interfacial bending energy [49]. At high salt concentrations, the viscosity could drop because of phase separation due to the salting out of surfactant.

Third, viscosity peaks are observed upon the addition of *amphiphilic molecules – cosurfactants*, typically fatty acids and alcohols, which are used as thickening agents [50-54]. In this case, the peak can be due again to transformation of the wormlike micelles into branched [52] or ribbonlike and disklike [53,55] aggregates. Alternatively, the peak could be related to the onset of a phase separation of surfactant as a precipitate from droplets and/or crystallites (see Section 3.4). Peaks in viscosity have been observed also in catanionic systems as a function of temperature [56] and in zwitterionic systems as a function of pH [57].

With respect to micelle topology, here we follow the terminology originating from Ref. [45], viz. the *entangled wormlike* micelle is a linear aggregate with two endcaps and no junctions with other micelles; a *branched* micelle consist of several connected branches, each of them beginning at a junction and ending with an endcap, and finally, the *multiconnected saturated network* represents a bicontinuous structure, where all endcaps have been transformed into intramicellar junctions.

In a preceding paper [58], we reported that the viscosity in mixed solutions of sulfonated methyl esters (SME) and cocamidopropyl betaine (CAPB) significantly increases with the rise of the total surfactant concentration. Our goal in the present article is to further extend this study and to examine the effects of surfactant composition, added salt and thickening agents, including the possible appearance of peaks in viscosity that could evidence synergistic growth, micelle shape transformations or phase separation.

The sulfonated methyl esters (SME) are produced from renewable palm-oil based materials [59-61] and have been promoted as alternatives to the petroleum-based surfactants [62]. SMEs exhibit a series of useful properties, such as excellent biodegradability and biocompatibility; excellent stability in hard water; good wetting and cleaning performance, and skin compatibility [60,63–69]. The SME surfactants are produced typically with even alkyl chainlengths, from $C_{12}$ to $C_{18}$. Here, they will be denoted $C_n$SME, $n$ = 12, 14, 16, 18. Information for the adsorption and micellar properties of $C_n$SME can be found in Refs. [70-72]. So far, there is a single study on the rheology of micellar SME solutions without cosurfactants (like CAPB) and thickeners, but in the presence of nanoparticles [73].

A complete systematic study on all possible combinations of the basic surfactants, cosurfactants (including fatty alcohols of various chainlengths) and salts exceeds the scope of the present article. Here, we focus on the most significant and interesting effects observed in our experiments with an emphasis on the interpretation of the obtained rheological data on the basis of theoretical models, in order to achieve a better understanding of the underlying molecular processes and phenomena.

In Section 2, we briefly describe the ingredients in the investigated systems and the used experimental methods. In Section 3, we report and systematize data from many rheological experiments in steady shear regime, which demonstrate the existence of strong synergism in the SME + CAPB system with respect to the rise of viscosity; effects of various additives on the synergistic maxima; salt curves with and without added fatty alcohol, and the effect of the concentration of thickeners – fatty alcohols and cocamide monoethanolamine (CMEA). Section 4 is dedicated to rheological experiments in oscillatory regime and to theoretical interpretation of the obtained results for the storage and loss moduli, $G'$ and $G''$. Part of the data exhibit standard rheological behavior and are interpreted in terms of the Maxwell viscoelastic model and the reptation-reaction model proposed by Cates [19] and developed in subsequent studies [20-25]. The latter model successfully explains both the Maxwellian behavior of micellar systems and the deviations from it. However, many of the investigated systems exhibit nonstandard rheological behavior. We have demonstrated that the rheological data for such systems can be theoretically described and analyzed in terms of an augmented version of the Maxwell model.

The paper could be useful for a broad audience of researchers, who are interested in synergistic effects in mixed micellar solutions and in their theoretical interpretation.

## 2. Materials and methods

*2.1. Materials*

In our study, we used two kinds of sulfonated methyl esters ($\alpha$-sulfo fatty acid methyl ester sulfonates, sodium salts, denoted also $\alpha$-MES), which are products of the Malaysian Palm Oil Board (MPOB) and KLK OLEO. The first one is myristic sulfonated methyl ester ($C_{14}$SME) 98%, $M$ = 344.34 g/mol, with critical micelle concentration CMC = 3.68 mM [70].

The second one, which will be denoted $C_{16,18}$SME, represents a mixture of 85 wt% palmitic ($C_{16}$SME) and 15 wt% stearic ($C_{18}$SME) sulfonated methyl esters and has a mean molecular weight $M$ = 376.70 g/mol and CMC = 0.9 mM. $C_{16,18}$SME is preferred in applications because of its better water solubility. Indeed, the Krafft temperatures of $C_{16}$SME and $C_{18}$SME are 28 and 40 °C, respectively, whereas their eutectic mixture $C_{16,18}$SME has 15 °C Krafft temperature and exhibits very good performance as detergent [74,75].

By conductometry, it was established that the used sulfonated methyl esters contain admixture of NaCl, which is 14 mol% for $C_{14}$SME and 24 mol% for $C_{16,18}$SME, relative to the surfactant [70]. To investigate the effect of salt on the micelle growth and viscosity rise, in some experiments we added NaCl, $M = 58.44$ g/mol, product of Sigma.

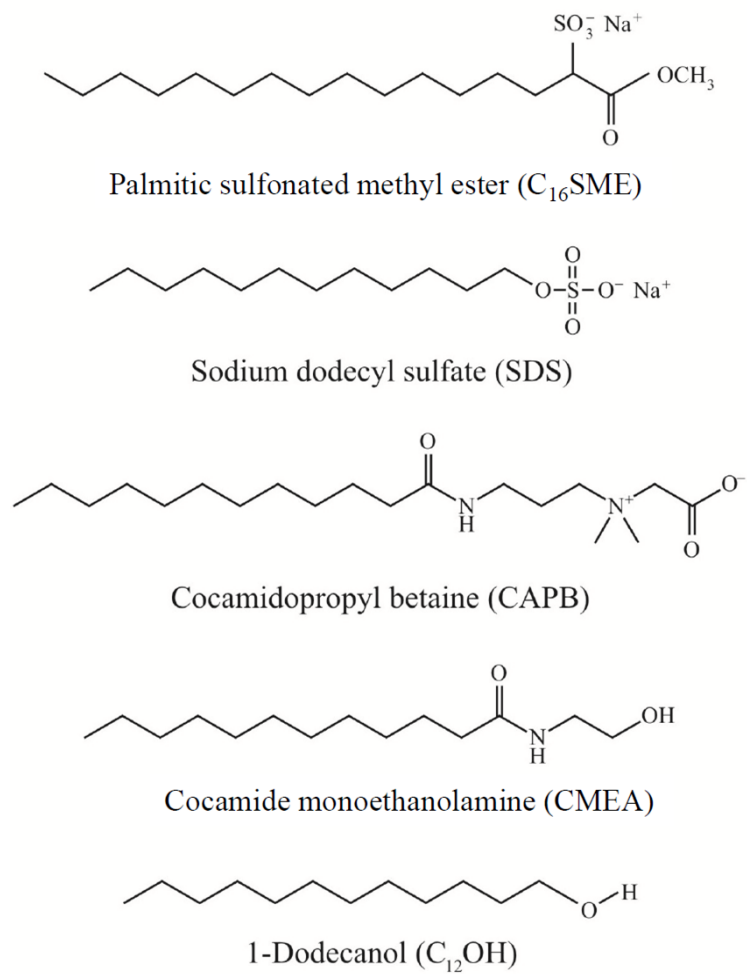

**Fig. 1.** Structural formulas of surfactant and co-surfactant molecules used in this study.

As a reference anionic surfactant (for comparison with the properties of SME), we used sodium dodecyl sulfate (SDS) 99%, $M = 288.38$ g/mol, product of Acros, with CMC = 8.0 mM.

We used also the zwitterionic surfactant cocamidopropyl betaine (CAPB), $M = 342.52$ g/mol, product of Evonik Nutrition & Care, GmbH Germany, with commercial name TEGO Betain F50; CMC = 0.089 mM. By conductometry, we established that 100 mM of the used CAPB contain an admixture of 118 mM NaCl. The structural formulas of surfactant and co-surfactant molecules used in this study are shown in Fig. 1.

As modifiers of the viscosity of the used concentrated micellar solutions (thickening and thinning agents), we used fatty alcohols: 1-decanol ($C_{10}OH$) 99%, $M$ = 158.28 g/mol, product of Aldrich, Cat. No 15,058-4; 1-dodecanol ($C_{12}OH$) 98%, $M$ = 186.34 g/mol, product of Sigma-Aldrich, Cat. No 12,659-9; 1-tetradecanol ($C_{14}OH$) 98%, $M$ = 214.39 g/mol, product of Merck, Cat. No S31409220, and 1-hexadecanol ($C_{16}OH$) 98%, $M$ = 242.48 g/mol, product of the Malaysian Palm Oil Board (MPOB). We investigated also rheological effects from the addition of the nonionic surfactant cocamide monoethanolamine (CMEA), $M$ = 206.14 g/mol, product of KLK OLEO and from the polymer guar hydroxypropyltrimonium chloride Jaguar C-13S, product of Solvay, which is used as oil-drop deposition agent in shampoo formulations [76].

All solutions were prepared with deionized water purified by Elix 3 water purification system (Millipore). The rheology modifiers (alcohols or polymer) were dissolved in the mixed micelles by vigorous agitation for one hour at 70 °C by using a temperature-controlled magnetic stirrer. The obtained multicomponent solutions were equilibrated in a thermostat at 30 °C for 24 h.

The working temperature was 30 °C. In separate experiments, the temperature was 25 °C, which is denoted on the respective graph. The chemicals were used as received without further purification. The natural pH of all studied solutions was between 5 and 6.

### 2.2. Experimental methods

The rheology of micellar solutions was measured by a rotational rheometer Bohlin Gemini (Malvern Instruments, UK) using cone-and-plate geometry. For less viscous samples of apparent viscosity $\eta \leq 40$ Pa·s, the cone angle was 2° and the minimal gap distance was 70 μm. The rheology experiments with more viscous solutions, $\eta > 40$ Pa·s, were carried out with cone angle 4° and gap distance 150 μm. The temperature, $T$ = 30 °C, was controlled by a Peltier element. The evaporation was suppressed by a solvent trap. To confirm the reproducibility of the data, each rheological experiment was repeated two times starting with newly prepared solutions.

The apparent viscosity $\eta \equiv \sigma / \dot{\gamma}$ (with $\sigma$ and $\dot{\gamma}$ being the experimental stress and share rate) was measured in *steady shear regime* as a function of $\dot{\gamma}$. The shear rate, $\dot{\gamma}$, was varied between 0.1 and 100 s$^{-1}$.

In addition, the storage and loss moduli, *G′* and *G″*, were measured in *oscillatory regime* as functions of the angular frequency, $\omega$, in the range from 0.06 to 100 rad/s at a small amplitude of the shear strain, viz. $\gamma_a = 0.02$.

In some of the investigated micellar solutions, a rise of turbidity due to crystallization was observed at the highest concentrations. To visualize and confirm the crystallites, Axioplan 2 microscope (Zeiss, Germany) was used in transmitted polarized light.

## 3. Rheology in steady shear regime

*3.1. Types of flow curves*

Fig. 2 illustrates the existence of two types of $\eta$-vs.-$\dot{\gamma}$ dependencies (flow curves): *regular* (Figs. 2a and b) and *irregular* (Figs. 2c and d). The experimental data are obtained with mixed solutions of $C_{14}$SME + CAPB and $C_{16,18}$SME + CAPB at two total surfactant concentrations, 8 and 12 wt%, in the presence of various additives: fatty alcohols, CMEA and NaCl. In our experiments, the weight fraction of CAPB, *w*, in the mixed surfactant solutions has been varied; *w* is defined as follows:

$$w = \frac{W_{\text{CAPB}}}{W_{\text{CAPB}} + W_{\text{anionic}}} \tag{1}$$

$W_{\text{CAPB}}$ and $W_{\text{anionic}}$ are the weights of CAPB and anionic surfactant (SDS or $C_n$SME) contained in the solution.

For the regular flow curves (Figs. 2a and b), the viscosity $\eta$ is constant at the lower shear rates $\dot{\gamma}$ (quasi-Newtonian behavior), whereas $\eta$ decreases almost linearly with $\dot{\gamma}^{-1}$ at the higher shear rates (shear thinning). As usual, the constant viscosity in the quasi-Newtonian (plateau) region will be termed zero-shear viscosity and denoted $\eta_0$. In this region, the apparent viscosity coincides with the real physical viscosity. Such shape of the flow curves is observed with both wormlike and branched micelles, and with polymer melts, as well.

For the systems with irregular behavior (Figs. 2c and d), no quasi-Newtonian regime is observed for most of the experimental curves in the investigated range of shear rates, $\dot{\gamma}$ (although plateau could be observed at lower $\dot{\gamma}$ using a more sensitive rheometer). For such experimental curves, zero-shear viscosity $\eta_0$ cannot be determined from the available data. Here, such curves are characterized with the value $\eta_{0.1}$ of $\eta$ at $\dot{\gamma} = 0.1$ s$^{-1}$ (see Section 3.4).

Note that the values of $\eta_0$ and $\eta_{0.1}$ can be comparable (Fig. 2). In the case of flow curves with plateau, $\eta_{0.1}$ coincides with $\eta_0$ if $\dot{\gamma} = 0.1$ s$^{-1}$ belongs to the plateau region.

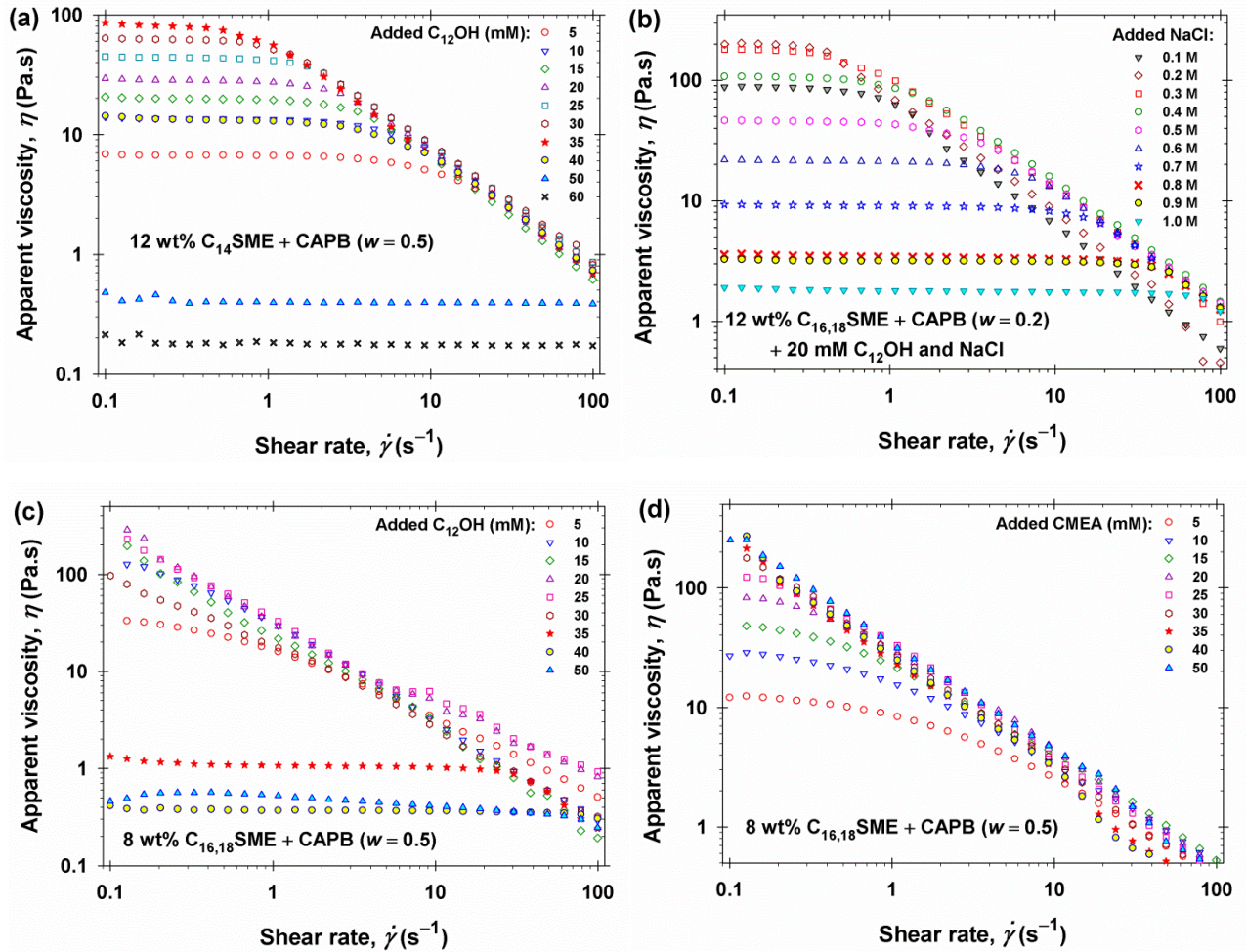

**Fig. 2.** Illustrative plots of the apparent viscosity, $\eta$, vs. the shear rate, $\dot{\gamma}$, for mixed solutions of C$_n$SME and CAPB at various weight fractions of CAPB, $w$, in the presence of different additives: (a) 12 wt% C$_{14}$CME + CAPB ($w$ = 0.5) at various concentrations of added dodecanol (C$_{12}$OH). (b) 12 wt% C$_{16,18}$CME + CAPB ($w$ = 0.2) + 20 mM C$_{12}$OH at various concentrations of added NaCl. (c) 8 wt% C$_{16,18}$CME ($w$ = 0.5) at various concentrations of added C$_{12}$OH. (d) 8 wt% C$_{16,18}$CME ($w$ = 0.5) at various concentrations of added cocamide monoethanolamine (CMEA).

Note that Fig. 2 has illustrative and auxiliary character – it visualizes the meaning of the quantities $\eta_0$ and $\eta_{0.1}$, for which experimental data are presented and analyzed in Section 3.2, beginning with binary and continuing with ternary mixtures.

*3.2. Effect of composition: anionic vs. zwitterionic surfactant*

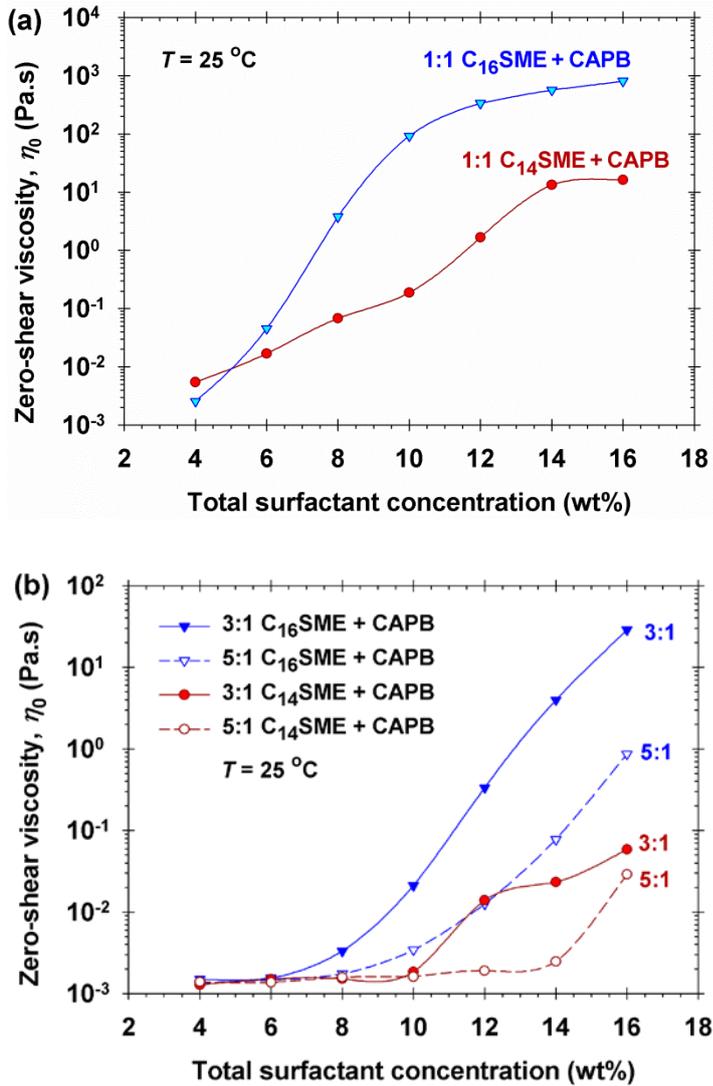

**Fig. 3**. Plots of experimental data from Ref. [58] for the zero-shear viscosity, $\eta_0$, vs. the total surfactant concentration for binary mixed solutions of $C_n$SME and CAPB ($n$ = 14, 16) at three different mass ratios denoted in the graphs: (a) 1:1 and (b) 3:1 and 5:1.

Fig. 3 presents experimental data for the zero-shear viscosity, $\eta_0$, vs. the total surfactant concentration for binary mixed solutions of $Cn$-SME and CAPB ($n$ = 14, 16) at three different fixed mass ratios, 1:1, 3:1 and 5:1. The experimental curves show that (i) the viscosity $\eta_0$ increases with the rise of total surfactant concentration reaching values up to $10^3$ Pa·s that indicate growth of giant micelles; (ii) $\eta_0$ is greater for the solutions with $C_{16}$SME as compared to $C_{14}$SME, and (iii) with the increase of the mass fraction of anionic surfactant (SME) from 1:1 to 5:1, the viscosity $\eta_0$ decreases.

Figs. 4a and b show plots of $\eta_0$ vs. $w$ for mixed solutions of an anionic surfactant (SDS and $C_{16,18}$SME) and a zwitterionic surfactant (CAPB) at a fixed total surfactant concentration of 12 wt%, without and with additives. As seen in these figures, the experimental curves have high maxima in viscosity, up to 665 Pa·s, at the intermediate values of $w$. In contrast, the values of $\eta_0$ are the lowest at the limiting points $w = 0$ and 1 (anionic and zwitterionic surfactant alone). To the best of our knowledge, for sulfonated methyl esters + CAPB the existence of such high maxima in viscosity is reported for the first time here.

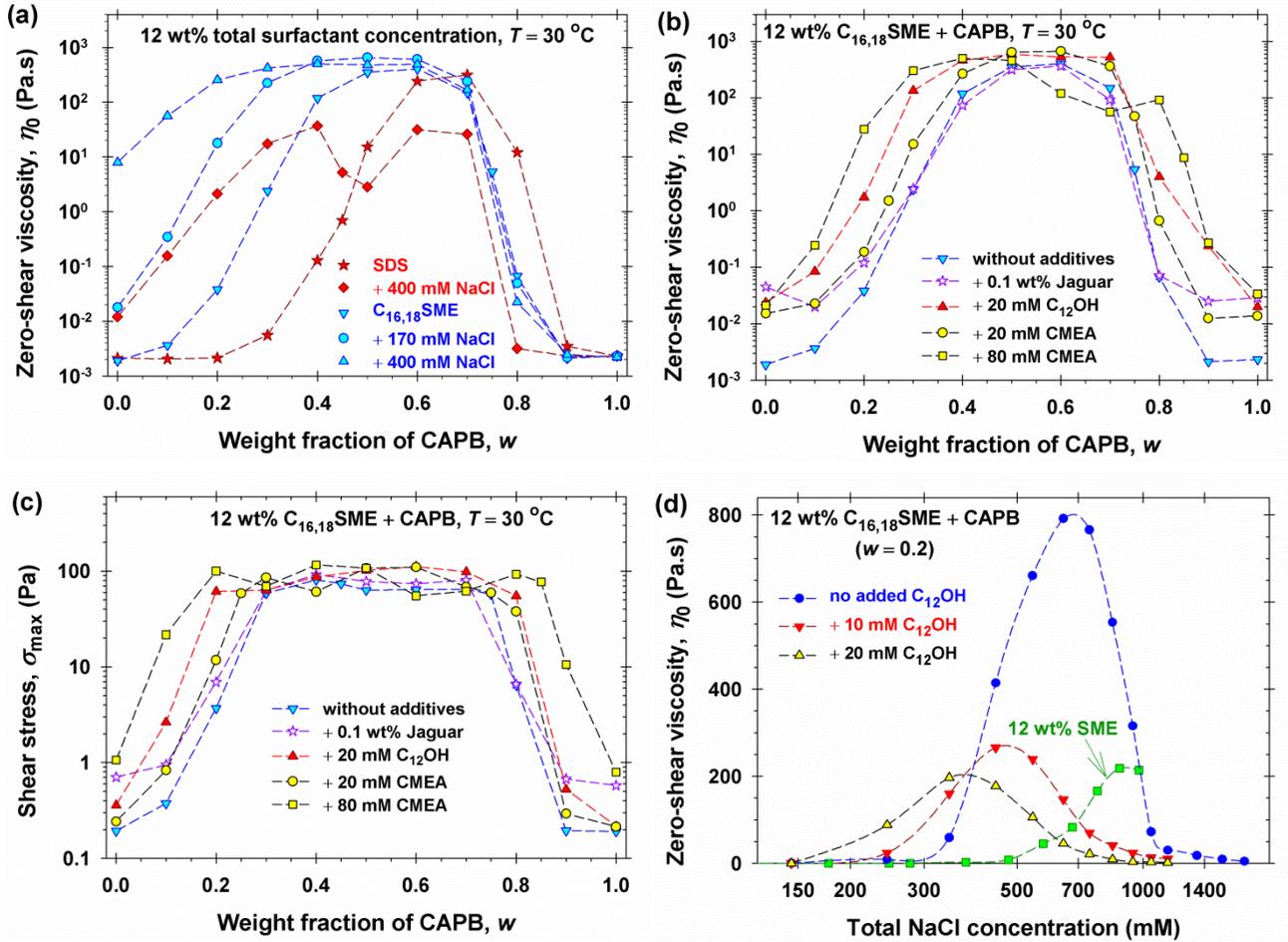

**Fig. 4**. Rheological curves with maxima: (a) Zero-shear viscosity, $\eta_0$, vs. the weight fraction of CAPB in the mixture with anionic surfactant (SDS or $C_{16,18}$SME), $w$, at 12 wt% total surfactant concentration in the presence of NaCl at various fixed concentrations. (b) and (c) Plots of $\eta_0$ vs. $w$ and $\sigma_{max}$ vs. $w$ for 12 wt% $C_{16,18}$SME + CAPB in the presence of various additives. (d) Salt curves: plots of $\eta_0$, vs. the total NaCl concentration for 12 wt% $C_{16,18}$SME + CAPB ($w = 0.2$) in the presence or absence of $C_{12}$OH.

The *separate* zwitterionic and anionic surfactants (without added salt) form small spherical or spheroidal micelles, whereas their mixing could lead to significant rise of viscosity; see e.g. [10]. The high peaks in viscosity in Figs. 4a and b can be explained with the growth of giant wormlike micelles due to synergistic interactions of the two kinds of surfactant. We could hypothesize that this effect is related to a favorable charge–dipole interaction between the headgroups of the anionic and zwitterionic surfactants. The initial increase of $\eta_0$ with the rise of $w$ (Figs. 4a and b) can be attributed to the increase of micelle length, whereas the decrease of $\eta_0$ to the right of the maximum – to micelle shortening. The fact that for all curves in Figs. 4a and b the maximal values of $\eta_0$ are of the same order of magnitude (despite of composition) could be interpreted as attainment of the limiting value of $\eta_0$ for very long micelles from such surfactants. (An exception with two maxima is the curve for 12 wt% SDS + 400 mM NaCl, which will be discussed below.)

In the presence of added NaCl, the synergistic peak broadens to the left (to lower $w$ values); see Fig. 4a. This effect is especially strong for $C_{16,18}$SME. Thus, at $w = 0.2$ (i.e. 4:1 $C_{16,18}$SME/CAPB) with added 400 mM NaCl the viscosity ($\eta_0 = 256$ Pa·s) is close to its maximal value. It is interesting, that the right branches of most viscosity curves (to the right of the maximum in Fig. 4a) are close to each other, irrespective of the NaCl concentration. This can be explained with the high contents of NaCl admixture in CAPB (see Section 2.1), which dominates the salt concentration at the higher values of $w$. Thus, for $w = 0.8$ (9.6 wt% CAPB) the concentration of $Na^+$ ions due to the surfactant is $\approx 400$ mM (330 mM from CAPB + 70 mM from $C_{16,18}$SME, including the salt admixture in $C_{16,18}$SME).

In Fig. 4a, the curve for SDS + CAPB + 400 mM added NaCl has two peaks. The left one could be explained with a shift of the maximum of the synergistic curve for SDS to the left due to the added NaCl. The right peak could be due to the increase of the total salt concentration at higher $w$ values, which is due to the high admixture of NaCl in CAPB.

Comparing the curves in Fig. 4a, we see that the viscosity maxima for $C_{16,18}$SME are higher and broader than those for SDS. Having in mind also the low sensitivity of $C_{16,18}$SME to hard water and its low skin irritation action, this makes the sulfonated methyl esters appropriate ingredients for shampoo formulations.

As a rule, the shampoo formulations contain also cationic polymers, like Jaguar C-13S, serving as deposition agents of oil drops to hair [76]. The viscosity curve with 0.1 wt% Jaguar C-13S in Fig. 4b indicates that the polymer dominates $\eta_0$ for $0 \leq w \leq 0.2$ and $0.8 \leq w \leq 1$ (at the lowest $\eta_0$ values), whereas in the region of the high synergistic maximum ($0.2 \leq w \leq 0.8$) the curves with and without polymer coincide. Hence, the effects of surfactants and Jaguar C-13S seem to be additive, i.e. there are no pronounced synergistic or antagonistic effects with respect to viscosity.

Addition of fatty alcohols also affects the viscosity of the concentrated micellar solutions of anionic and zwitterionic surfactants. As seen in Fig. 4b, the addition of 20 mM dodecanol ($C_{12}OH$) to the solutions with 12 wt% $C_{16,18}SME$ + CAPB markedly increases $\eta_0$ for almost all $w$ values. For example, at $w = 0.25$ (3:1 $C_{16,18}SME$ + CAPB), the viscosity increases from 0.304 to 15.75 Pa·s upon the addition of 20 mM (0.37 wt%) $C_{12}OH$, that is 52 times increase of $\eta_0$. In contrast, in Fig. 3b for the curve with 12 wt% 3:1 $C_{16,18}SME$ + CAPB, the increase of the total surfactant concentration with 0.37 wt% (from 12 to 12.37 wt%) leads only to 5 % increase of $\eta_0$. Note, however, that the effect of dodecanol depends on the concentration of added salt (see below).

The effect of 20 mM CMEA on viscosity turns out to be somewhat weaker than that of 20 mM $C_{12}OH$; see Fig. 4b. In other words, CMEA turns out to be a weaker thickening agent for such type of micellar solutions. Note, however, that the highest viscosities in the same figure are obtained by the addition of 80 mM CMEA. On the other hand, the solutions with 80 mM added $C_{12}OH$ are phase separated (turbid) and have a lower viscosity (see Section 3.4). Hence, CMEA can be used as a thickening agent in a wider concentration range than $C_{12}OH$, so that by addition of CMEA one could reach higher viscosity values.

The plot of the measured macroscopic shear stress, $\sigma \equiv \eta\dot{\gamma}$, vs. the shear rate, $\dot{\gamma}$, has a maximum as observed in other studies; see e.g. Ref. [52] – the supporting information therein. Here, the maximal shear stress is denoted $\sigma_{max}$. In Fig. 4c, the plots of experimental data for $\sigma_{max}$ vs. $w$ resemble (by shape) the plots of $\eta_0$ vs. $w$ for the same systems: $\sigma_{max}$ has its lower values at $w = 0$ and 1 (for the separate surfactants), whereas for $0.2 < w < 0.8$ exhibits a high and flat maximum of height 70–100 Pa, which seems to be the limiting range of values of $\sigma_{max}$ for very long micelles. The shape of the plot $\sigma_{max}$ vs. $w$ is another manifestation of the synergism upon mixing of SME with CAPB.

*3.3. Salt curves – effect of NaCl concentration*

In Fig. 4d, the surfactant composition is fixed ($w = 0.2$), whereas the NaCl concentration is varied. Along the abscissa, the total NaCl concentration is plotted, which includes both the added salt and the salt present as an admixture in the surfactants; see Section 2.1. Each experimental dependence has a typical shape of curve with maximum, called the salt curve [37-42]. In Fig. 4d, the curve with the highest maximum (ca. 800 Pa·s) is that for 12 wt% $C_{16,18}SME$ + CAPB without additives. For 12 wt% $C_{16,18}SME$ alone (no CAPB), the viscosity also increases with the rise of salt concentration, but its maximal value is smaller (218 Pa·s); at ca. 1000 mM NaCl (where this experimental curve is interrupted) the solutions become turbid, which indicates phase separation due to salting out of $C_{16,18}$SME. The data show that the presence of CAPB considerably increases the height of the salt curve. The comparison of the maximal measured values of $\eta_0$ for the single component system (218 Pa·s) and binary mixture (792 Pa·s at $w = 0.2$) shows 3.6 times increase of viscosity in the presence of 20 % CAPB.

In contrast, the addition of fatty alcohol (dodecanol) markedly decreases the height of the salt curves – see the curves with 10 and 20 mM $C_{12}$OH in Fig. 4d. Note that these two curves intersect the salt curve without $C_{12}$OH. Thus, to the left of the intersection point (below ca. 390 mM NaCl) $C_{12}$OH acts as a thickening agent, whereas to the right of the intersection point (above 390 mM NaCl) $C_{12}$OH acts as a thinning agent.

As already mentioned, the rise of viscosity with the increase of salt concentration (the left branch of the salt curves) is probably related to the growth and entanglement of wormlike micelles, whereas the subsequent decrease of viscosity (the right branch) could be explained with a transition to branched micelles [44-48]. At the highest salt concentrations, one could observe phase separation due to the salting out of surfactant. In practical applications, the variation of salt concentration is used for tuning the viscosity of shampoos (and other similar) formulations.

*3.4. Effect of added fatty alcohols and CMEA*

Additional data for the effect of added fatty alcohols on the viscosity of mixed $C_n$SME + CAPB solutions are shown in Fig. 5, where results for the effect of CMEA are also presented. Note that the flow curves Figs. 5a and b are like those shown in Figs. 2b and c, respectively, and for this reason they are characterized by $\eta_0$ and $\eta_{0.1}$.

The data show that the increase of concentration of added fatty alcohol ($C_n$OH, $n = 10$, 12, 14, 16) up to 30–35 mM leads to a significant increase of viscosity. In other words, in this concentration range the fatty alcohols act as thickeners, supposedly, they promote the growth of longer wormlike micelles. For example, in the case of 12 wt% 1:1 $C_{14}$SME + CAPB, the viscosity $\eta_0$ increases from 1.8 to 109 Pa·s upon the addition of tetradecanol, $C_{14}$OH (Fig. 5a). Similarly, in the case of 8 wt% 1:1 $C_{16,18}$SME + CAPB, the viscosity $\eta_{0.1}$ increases from 3.8 to 291 Pa·s (at the maximum) upon the addition of $C_{14}$OH, (Fig. 5b). The effects of $C_{10}$OH, $C_{12}$OH and $C_{14}$OH on viscosity are comparable, whereas the effect of $C_{16}$OH is markedly weaker.

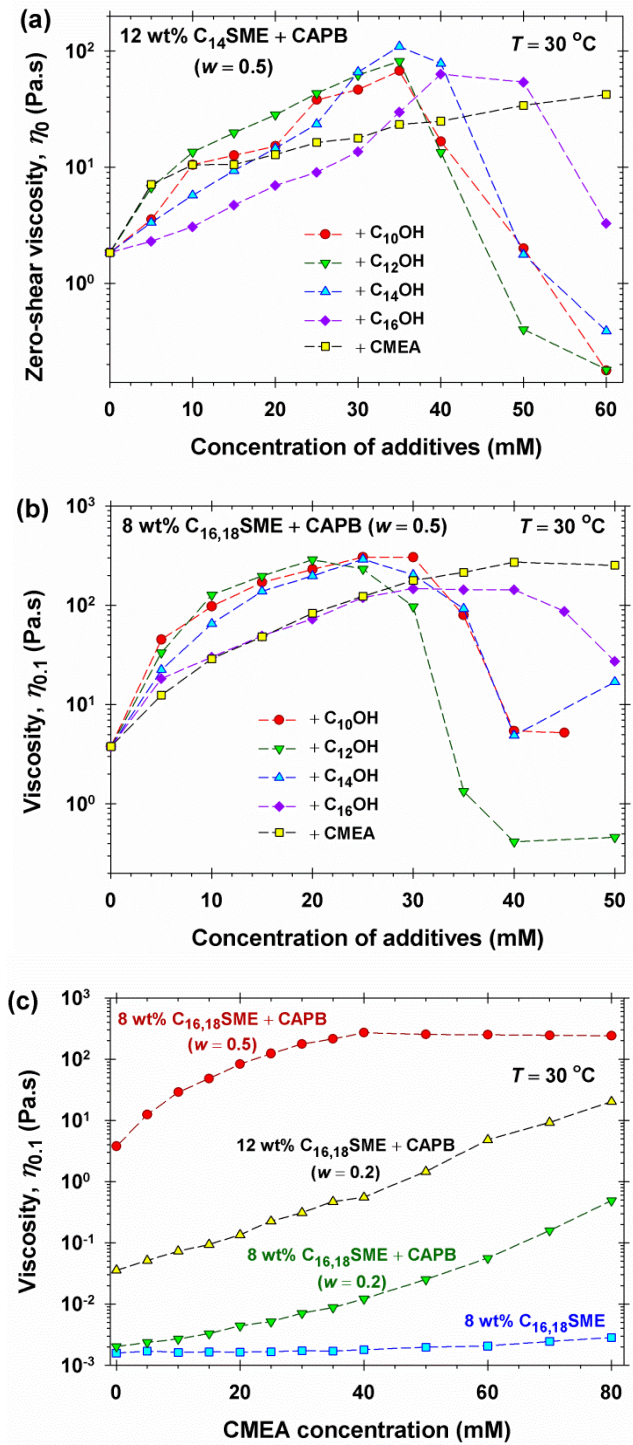

**Fig. 5**. Effect of additives on the viscosity of mixed $C_n$-SME + CAPB solutions: (a) $\eta_0$ vs. the concentration of additive ($C_n$OH or CMEA) for a basic solution of 12 wt% $C_{14}$SME + CAPB ($w = 0.5$). (b) $\eta_{0.1}$ vs. the concentration of additive for a basic solution of 8 wt% $C_{16,18}$SME + CAPB ($w = 0.5$). (c) $\eta_{0.1}$ vs. the concentration of CMEA for basic solutions of $C_{16,18}$SME + CAPB of different compositions denoted in the figure.

To the right of the maxima of the experimental curves with $C_n$OH in Figs. 5a and b, the solutions become turbid because of the precipitation of small droplets or crystallites (depending on the system), which are observable by optical microscope. At that, solution's viscosity decreases. The latter indicates that the precipitate contains both the basic surfactants (SME and/or CAPB) and fatty alcohol. Indeed, if the precipitation was just a solubility limit of the fatty alcohol in the wormlike micelles, as in Ref. [77], and crystals of the excess $C_n$OH appear on the background of the formed entangled wormlike micelles, the viscosity should increase, rather than decrease. In other words, the precipitation is related to transformation of the wormlike micelles into drops or crystallites.

The thickening effect of CMEA is comparable with those of the fatty alcohols, with the only difference that the viscosity monotonically increases with the addition of CMEA in the studied concentration range (up to 80 mM CMEA) and no precipitation was observed (Fig. 5). The effects of CMEA on different systems are compared in Fig. 5c. In this figure, the lowest experimental curve shows that the addition of CMEA to 8 wt% $C_{16,18}$SME produces a rather weak effect on viscosity. In the case of mixed solutions of 8 and 12 wt% $C_{16,18}$SME + CAPB at $w = 0.2$, the addition of CMEA leads to a stable growth of viscosity. Finally, for the upper curve in Fig. 5c, at 8 wt% $C_{16,18}$SME + CAPB, $w = 0.5$, the effect of CMEA exhibits a saturation at concentrations $\geq 40$ mM CMEA, where $\eta_{0.1} \approx 250$ Pa·s, which probably corresponds to the limiting viscosity for very long micelles.

In summary, in a wide concentration range, $C_n$OH and CMEA can serve as thickening agents for concentrated mixed solutions of $C_n$SME and CAPB (Fig. 4).

**4. Rheology in oscillatory regime**

*4.1. Experimental results for G' and G''; comparison with the Maxwell model*

In the case of experiments with rotational rheometer in oscillatory regime, sinusoidal oscillations of the strain are imposed:

$$\gamma(t) = \gamma_a \sin(\omega t) \tag{2}$$

where $\gamma_a$ is amplitude, $t$ is time and $\omega$ is angular frequency. Our experiments were carried out at $\gamma_a = 0.02$, whereas $\omega$ was varied. As a rule, the measured stress, $\sigma(t)$, is phase-shifted, so that it can be expressed in the form:

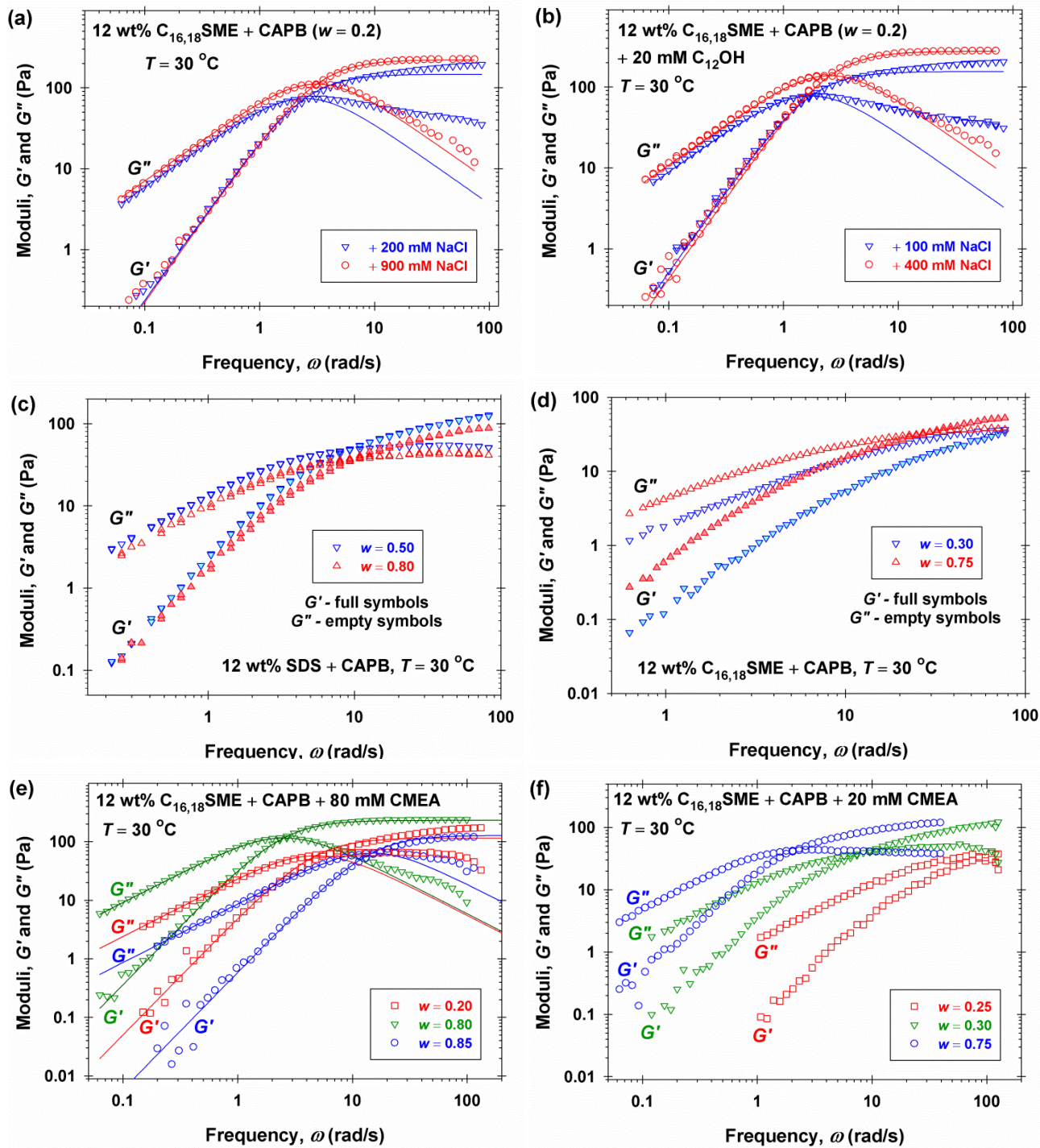

**Fig. 6.** Comparison of systems with standard (a,b,e) and nonstandard (c,d,f) rheological behavior: Illustrative plots of the storage and loss moduli, $G'$ and $G''$, vs. frequency $\omega$. (a) 12 wt% $C_{16,18}$SME + CAPB ($w = 0.2$) at 200 and 900 mM added NaCl; (b) 12 wt% $C_{16,18}$SME + CAPB ($w = 0.2$) + 20 mM $C_{12}$OH at 100 and 400 mM added NaCl; (c) 12 wt% SDS + CAPB at $w = 0.50$ and 0.80; (d) 12 wt% $C_{16,18}$SME + CAPB at $w = 0.50$ and 0.75; (e) 12 wt% $C_{16,18}$SME + CAPB + 80 mM CMEA at $w = 0.20$, 0.80 and 0.85; (f) 12 wt% $C_{16,18}$SME + CAPB + 20 mM CMEA at $w = 0.25$, 0.30 and 0.75. The solid lines represent best fits with the standard Maxwell model, Eq. (4) of data at lower $\omega$ values.

$$\frac{\sigma(t)}{\gamma_a} = G' \sin(\omega t) + G'' \cos(\omega t) \tag{3}$$

where $G'$ and $G''$ are the storage and loss moduli, respectively; $G'$ and $G''$ are independent of $t$, but they depend on $\omega$; see Fig. 6. In Eq. (3), it is presumed that both the amplitude $\gamma_a$ and frequency $\omega$ are small enough, so that the contributions of higher-order Fourier modes to $\sigma(t)$ are negligible, i.e. we are dealing with *linear* rheological response.

The obtained $G'(\omega)$ and $G''(\omega)$ experimental curves, like those in Fig. 6, are usually compared with the predictions of the Maxwell model of viscoelastic body [78]:

$$G'(\omega) = G_0 \frac{(\omega \tau_R)^2}{1+(\omega \tau_R)^2}, \quad G''(\omega) = G_0 \frac{\omega \tau_R}{1+(\omega \tau_R)^2} \tag{4}$$

See Section 4.2 for more details; $G_0$ and $\tau_R$ are constants, which are, respectively, the characteristic shear elastic modulus and relaxation time of the Maxwell model. The basic properties of the $G'(\omega)$ and $G''(\omega)$ dependencies, as given by Eq. (4) are as follows: (i) $G''(\omega)$ has a maximum at $\omega \tau_R = 1$, and (ii) the $G'(\omega)$ and $G''(\omega)$ curves intersect exactly in the point of the maximum of

$$G''(\omega): \quad G'(\omega_c) = G''(\omega_c) = \frac{G_0}{2}; \quad \omega_c \equiv \frac{1}{\tau_R} \tag{5}$$

where $\omega_c$ is the *crossover* frequency.

Comparing the shape of the experimental $G'(\omega)$ and $G''(\omega)$ dependences with the predictions of the Maxwell model, Eq. (4), one could distinguish two types of rheological behavior; see also Refs. [22,79,80]. (i) For the systems with *standard* rheological behavior, the experimental $G'(\omega)$ and $G''(\omega)$ curves comply with the Maxwell model up to the crossover point, $0 \leq \omega \leq \omega_c$, whereas there can be deviations for $\omega \geq \omega_c$. (ii) For the systems with *nonstandard* rheological behavior, the experimental $G'(\omega)$ and $G''(\omega)$ curves either markedly deviate from the Maxwell model at frequencies below the crossover point, or crossover point is not observed at all in the investigated frequency range.

All experimental dependences in Fig. 6 are obtained with systems, for which the flow curves in steady-shear regime are regular, like those in Figs. 2a and b. For all of them the total concentration of the two basic surfactants (anionic and zwitterionic) is the same, viz. 12 wt%.

In oscillatory regime, standard rheological behavior is observed with 12 wt% $C_{16,18}$SME + CAPB ($w = 0.2$) in the presence of added salt (Figs. 6a and b), or in the presence of added 80 mM CMEA (Fig. 6e).

Nonstandard rheological behavior is observed for 12 wt% SDS + CAPB and 12 wt% $C_{16,18}$SME + CAPB without additives (Figs. 6c and d), as well as for 12 wt% $C_{16,18}$SME + CAPB with added 20 mM CMEA (Fig. 6f).

*4.2. Theoretical background and data interpretation*

From the above experimental data (Section 4.1 and other similar data) one could obtain information for the micellar properties and processes. To do that, we have to compare the data with the predictions of available theoretical models and determine the values of basic descriptive parameters, which could be the phenomenological parameters $G_0$ and $\tau_R$ of the Maxwell viscoelastic model, and the micellar reptation and breakage times, $\tau_{rep}$ and $\tau_{br}$, of the Cates reptation-reaction model [19,25]. Here, we first briefly recall the logical scheme of the standard Maxwell model, in view of its further generalization in Section 4.4 for describing systems with nonstandard rheological behavior. The emphasis in our considerations will be on the procedures for comparison of theory and experiment, and on the interpretation of the obtained parameter values.

The Maxwell model with sequentially connected elastic element of elastic and viscous elements is illustrated in Fig. 7a. The stress $\sigma$ exerted on the two elements is the same, whereas the strains (and strain rates) of the two elements are additive: $\dot{\gamma} = \dot{\gamma}_e + \dot{\gamma}_v = \dot{\sigma}/G + \sigma/\eta$, where $G$ and $\eta$ are the characteristic elasticity and viscosity. In this way, we arrive to the basic equation of the Maxwell model:

$$\frac{d\sigma}{dt} + \nu_{ch}\sigma = G\frac{d\gamma}{dt}, \quad \nu_{ch} \equiv \frac{G}{\eta} \tag{6}$$

where $\nu_{ch}$ is a characteristic frequency of the system. In the standard Maxwell model, $G$ and $\eta$ are constant:

$$G \equiv G_0, \quad \eta \equiv \eta_0, \quad \nu_{ch} \equiv \frac{G_0}{\eta_0} \equiv \frac{1}{\tau_R} \tag{7}$$

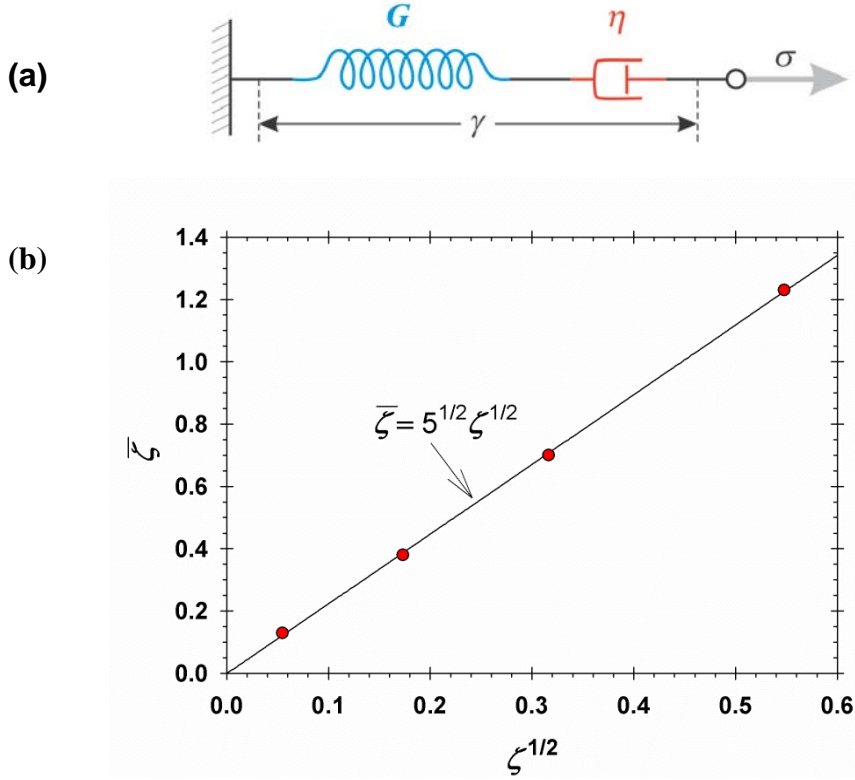

**Fig. 7.** (a) The Maxwell viscoelastic model with sequentially connected spring of elasticity $G$ and dash-pot with viscosity $\eta$. (b) The reptation-reaction model: plot $\bar{\zeta}$ vs. $\zeta^{1/2}$ with data from Ref. [23], where $\bar{\zeta} = \tau_R/\tau_{rep}$ and $\zeta = \tau_{br}/\tau_{rep}$; $\tau_R$, $\tau_{br}$ and $\tau_{rep}$ are the characteristic relaxation, breakage and reptation times, respectively. The solid line is a linear regression.

(In Section 4.4, $\omega$-dependent $G$ and $\eta$ will be considered.) If the body is subjected to deformation, which is then fixed ($\dot{\gamma} = 0$), Eq. (6) describes an exponential stress relaxation process, $\sigma(t) \propto \exp(-t/\tau_R)$. Physically, the stress accumulated in the elastic element is gradually dissipated in the viscous element; see Fig. 7a. The fact that the Maxwell model is related to a single exponential relaxation process is important for understanding the reasons for deviations from this model [22]; see below. Furthermore, substituting Eqs. (2) and (3) into Eq. (7), and setting equal the coefficients before the sine and cosine, we obtain Eq. (4) with $\tau_R = \eta_0/G_0$. In view of the last equation, the two relations in Eq. (4) can be solved with respect to $G_0$ and $\eta_0$:

$$G_0 = \frac{G'^2 + G''^2}{G'}, \qquad \eta_0 = \frac{G'^2 + G''^2}{G''\omega} \tag{8}$$

Further, it is important to note that $G'$ and $G''$ satisfy an equation of circle of radius $G_0/2$:

$$(G' - G_0/2)^2 + (G'')^2 = (G_0/2)^2 \tag{9}$$

Substituting Eq. (4) into Eq. (9), one can check the validity of the latter equation.

In Figs. 8a and b, we have plotted data for $G'$ and $G''$ for various systems with standard rheological behavior in accordance with Eq. (9) – this is the so called the Cole-Cole plot [23,24]. (Such plot was first introduced by Cole and Cole [81] for presenting data for the frequency dependence of dielectric constant of polar liquids.) $G_0$ is determined as $2G'$ in the crossover point; see Eq. (5). In Figs. 8a and b, the semicircle of radius 1 represents the prediction of the Maxwell model. For $G'/(G_0/2) > 1$, which is the region of higher frequencies, $\omega > \omega_c$, the experimental curves deviate from the Maxwell model.

Theoretical interpretation of the rheological behavior of surfactant solutions with giant micelles was given in the framework of the reptation-reaction model by Cates [19], which was extended in subsequent studies [20-25]. The process of reptation is characterized by the respective relaxation time, $\tau_{rep}$, and the reaction of micelle reversible breakage and recombination is characterized by another relaxation time, $\tau_{br}$. The reptation-reaction model predicts that for sufficiently small ratio $\tau_{br}/\tau_{rep}$ (and not too high frequencies, $\omega$) the stress relaxation in the system is exponential and is characterized by a single relaxation time $\tau_R \propto (\tau_{rep}\tau_{br})^{1/2}$. This result explains the fact that the experimental data obey the Maxwell model for not too high frequencies – for $G'/(G_0/2) < 1$ in Figs. 8a and b.

At higher frequencies, $\omega$, the reptation-reaction model predicts deviations from the Maxwell model, which are presented by the solid lines to the right of the semicircles in Figs. 5a and b. This prediction is also in good agreement with the experimental results. In this case, there are no explicit analytical expressions for $G'$ and $G''$, and only numerical solutions are available. From a theoretical viewpoint, the predicted (and observed) deviations from the Maxwell model at $\omega > \omega_c$ are due to a transition from single-exponential relaxation to multi-exponential or nonexponential relaxation processes [25].

The comparison of the reptation-reaction model with experimental data allows one to characterize the investigated system with physical parameters, such as the characteristic micelle breakage and reptation times, $\tau_{br}$ and $\tau_{rep}$. For this goal, let us consider the parameters $\zeta$ and $\bar{\zeta}$ introduces in Refs. [19,25]:

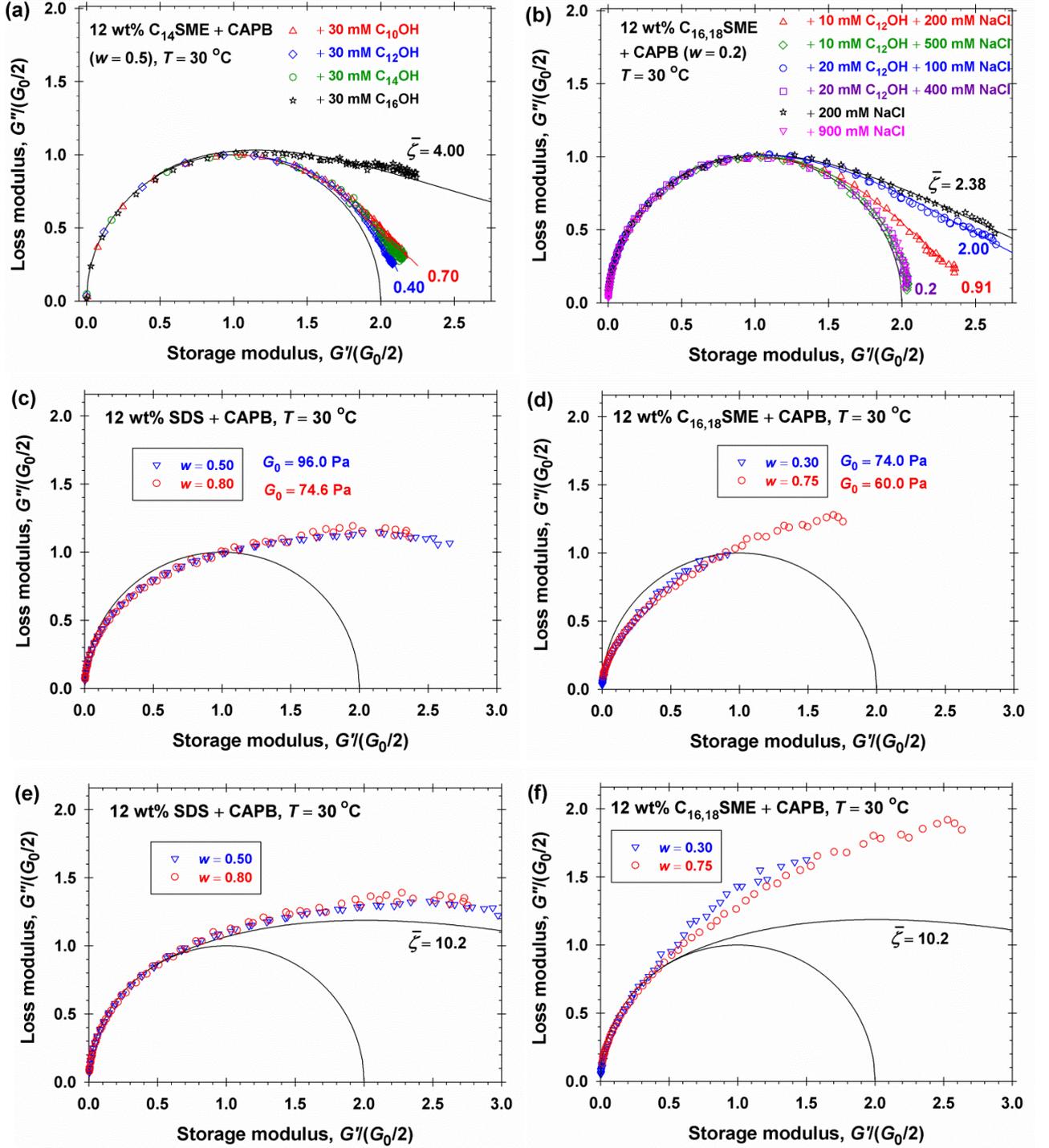

**Fig. 8.** Comparison of Cole-Cole plots for systems with standard (a,b) and nonstandard (c,d,e,f) rheological behavior. (a,b) Systems with standard behavior, for which the semicircle is drawn with $G_0 = 2G'$ at the crossover point; at higher $G'$, the values of $\bar{\zeta}$ are determined by using the reptation-reaction model (see the text). (c,d) Systems with nonstandard behavior: the semicircle is drawn again with $G_0 = 2G'$ at the crossover point, but the agreement with the experimental points is poor. (e,f) For the same systems, $G_0$ can be determined by data fit with Eq. (9) at the lower $G'$ values; for comparison, the upper solid line from Ref. [24] shows the prediction of the reptation-reaction model for $\bar{\zeta} = 10.2$.

$$\zeta \equiv \frac{\tau_{\mathrm{br}}}{\tau_{\mathrm{rep}}}, \quad \bar{\zeta} \equiv \frac{\tau_{\mathrm{br}}}{\tau_{\mathrm{R}}} \tag{10}$$

The theoretical dependence $\bar{\zeta}(\zeta)$ can be obtained only by numerical solutions of the respective statistical equations. To quantify this dependence, following Anachkov et al. [53], in Fig. 7b we have plotted numerical data for $\zeta$ and $\bar{\zeta}$ from Ref. [23]. As seen, the plot excellently agrees with a straight line, which corresponds to the relation:

$$\zeta = \frac{\bar{\zeta}^2}{5} \tag{11}$$

Combining Eqs. (10) and (11), one can estimate the proportionality coefficient in the relation $\tau_{\mathrm{R}} \propto (\tau_{\mathrm{rep}} \tau_{\mathrm{br}})^{1/2}$, viz.

$$\tau_{\mathrm{R}} = \frac{1}{\sqrt{5}} (\tau_{\mathrm{rep}} \tau_{\mathrm{br}})^{1/2} \approx 0.447 (\tau_{\mathrm{rep}} \tau_{\mathrm{br}})^{1/2} \tag{12}$$

In the mean field theory [19], the relation $\tau_{\mathrm{R}} \propto (\tau_{\mathrm{rep}} \tau_{\mathrm{br}})^{1/2}$ was derived analytically assuming $\zeta \leq 1$. However, the results of numerical calculations reported in the same study indicate that the relation $\tau_{\mathrm{R}} \propto (\tau_{\mathrm{rep}} \tau_{\mathrm{br}})^{1/2}$ and the related Eq. (11) have a wider validity. Indeed, the double log plot of the dimensionless viscosity $f$ vs. $\zeta$ in Ref. [19] shows that the linear portion of the plot (corresponding to $f \propto \zeta^{1/2}$) spans up to $\zeta \approx 10$. Consequently, the interval of applicability of Eqs. (11) and (12) can be expanded to $\zeta \leq 10$, which corresponds to $\bar{\zeta} \leq \sqrt{50} \approx 7$. This applicability limit of Eq. (11) was used in our estimates of $\zeta$ and $\tau_{\mathrm{rep}}$ (Sections 4.3 and 4.6).

For flexible micelles in semi-diluted regime, knowing $G_0$ one can estimate the mesh size (the correlation length) of the transient micellar network $\xi$ [18,82]:

$$\xi \approx \left( \frac{k_{\mathrm{B}} T}{G_0} \right)^{1/3} \tag{13}$$

where $k_{\mathrm{B}}$ is the Boltzmann constant and $T$ is the absolute temperature. In addition, we have $\xi \propto L_{\mathrm{p}}^{2/5} L_{\mathrm{e}}^{3/5}$, where $L_{\mathrm{p}}$ is the persistence length, at which the thermal motion overcomes the bending rigidity and the micelle resembles a flexible polymer chain; $L_{\mathrm{e}}$ is the entanglement length, i.e., the contour length along a wormlike micelle between two entanglement points [18,25,28,82,83]. Thus, one obtains the following estimate for $G_0$:

$$G_0 \propto k_B T L_p^{-6/5} L_e^{-9/5} \tag{14}$$

The characteristic breakage time, $\tau_{br}$, which is the survival time of a chain of mean length $\bar{L}$ before it breaks into two pieces, can be expressed as follows [19,22]:

$$\tau_{br} = \frac{1}{k_1 \bar{L}} \propto \varphi^{-0.5} \tag{15}$$

where $k_1$ is the rate constant of breakage; $\varphi$ is the surfactant volume fraction, and we have used the fact that $\bar{L} \propto \varphi^{0.5}$ [84]. Eq. (15) implies that $\tau_{br}$ is smaller for longer micelles, for which the probability to break is greater. Moreover, $\tau_{br}$ should be smaller for mechanically labile micelles, for which $k_1$ greater. Other useful relations for semidilute solutions of wormlike micelles are [18,22,26]:

$$\xi \propto \varphi^{-0.77}; \quad G_0 \approx k_B T \xi^{-3} \propto k_B T \varphi^{2.3} \tag{16}$$

The combination of the above relations yields:

$$\tau_{rep} \propto \bar{L}^3 \xi^{-6} \varphi^{-3} \propto \varphi^{3.1} \tag{17}$$

Following Ref. [26], for the model with reversible scission we obtain:

$$\tau_R \propto (\tau_{br}\tau_{rep})^{1/2} \propto \bar{L} \xi^{-3} \varphi^{-1.5} \propto \varphi^{1.3} \tag{18}$$

$$\eta_0 = G_0 \tau_R \propto \bar{L} \xi^{-6} \varphi^{-1.5} \propto \varphi^{3.6} \tag{19}$$

Hence, only $\tau_{br}$ decreases with the rise of total surfactant volume fraction, $\varphi$, whereas $\eta_0$, $\tau_{rep}$, $G_0$ and $\tau_R$ increase with $\varphi$. At *fixed* total surfactant concentration $\varphi$ (but variable composition of the mixed solution), the micelle length $\bar{L}$ could grow owing to the addition of salt or cosurfactant. Then $\tau_{br}$ diminishes $\propto \bar{L}^{-1}$; $\tau_R$ and $\eta_0$ should grow $\propto \bar{L}$, whereas $\tau_{rep}$ should strongly increase $\propto \bar{L}^3$; $G_0$ might vary if the changes in composition affect $L_p$ and $L_e$; see Eq. (14). In the next subsection, we verify how these theoretical predictions compare with the experiment.

*4.3. Systems with standard rheological behavior: numerical results*

The procedure for data processing is as follows. We dispose with sets of experimental data for $G'(\omega)$ and $G''(\omega)$ like those in Figs. 6a, b and e.

(1) From the coordinates of the crossover point, we determine $\tau_R = 1/\omega_c$ and $G_0 = 2G'(\omega_c) = 2G''(\omega_c)$; see Eq. (5). Alternatively, $G_0$ can be determined from the fit of the Cole-Cole plot ($G'$ vs. $G''$) at the smaller $G'$ values with circle in accordance with Eq. (9).

(2) The experimental Cole-Cole plot, scaled with $G_0/2$ as in Fig. 8, is compared with available theoretical Cole-Cole plots [23,24,48], which are calculated by the reptation-reaction model for a given $\bar{\zeta}$, and are scaled to unit circle. The comparison yields the value of $\bar{\zeta}$ for the respective experimental curve. The accuracy of determination of $\bar{\zeta}$ could be increased by appropriate linear interpolation.

(3) Finally, from the determined $\bar{\zeta}$ one calculates $\tau_{br} = \bar{\zeta}\tau_R$; $\zeta = \bar{\zeta}^2/5$, and $\tau_{rep} = \tau_{br}/\zeta$; see Eqs. (10) and (11).

**Table 1.** Rheological parameters for the systems of standard rheological behavior in Fig. 8a: 12 wt% $C_{14}$SME + CAPB ($w = 0.5$) + 30 mM $C_n$OH, $n = 10, 12, 14,$ and 16.

| $C_n$OH | $\eta_0$ (Pa·s)* | $G_0$ (Pa) | $\tau_R$ (s) | $G_0\tau_R$ (Pa·s) | $\xi$ (nm) | $\bar{\zeta}$ | $\zeta$ | $\tau_{br}$ (s) | $\tau_{rep}$ (s) |
|---|---|---|---|---|---|---|---|---|---|
| $C_{10}$OH | 46.4 | 194 | 0.229 | 44.5 | 27.8 | 0.70 | 0.098 | 0.161 | 1.64 |
| $C_{12}$OH | 62.1 | 216 | 0.280 | 60.4 | 26.9 | 0.40 | 0.032 | 0.112 | 3.50 |
| $C_{14}$OH | 66.0 | 190 | 0.344 | 65.3 | 28.0 | 0.70 | 0.098 | 0.241 | 2.46 |
| $C_{16}$OH | 13.6 | 98.0 | 0.140 | 13.7 | 35.0 | 4.00 | 3.20 | 0.560 | 0.175 |

*The value of $\eta_0$ is taken from the plateau of the respective flow curve, like those in Fig. 2a.

Table 1 shows results obtained from the experimental curves in Fig. 8a, which differ by the chainlengths of the added fatty alcohols, $C_n$OH, $n = 10, 12, 14,$ and 16. It is remarkable that the values of $G_0\tau_R$ determined from the data in oscillatory regime are in excellent agreement with the values of $\eta_0$ *independently* obtained from the plateaus of the respective flow curves, like those in Fig. 2a; see Eq. (7). Such coincidence is sometimes cited as fulfillment of the Cox-Merz rule [85]. This agreement is an argument in favor of the correctness and self-consistence of the used procedure for comparison of theory and experiment [52,53].

Because the total surfactant concentration is fixed (Table 1), among the calculated parameters $\tau_{rep} \propto \bar{L}^3$ is the most sensitive to the length of the formed micelles; see Eqs. (15)–(19). Hence, in the investigated $C_{14}$SME + CAPB system the dodecanol ($C_{12}$OH) is the strongest promoter of micelle growth, followed by the tetradecanol ($C_{14}$OH), whereas the effect of hexadecanol ($C_{16}$OH) is the weakest. Correspondingly, the breakage time $\tau_{br}$ is the shortest for $C_{12}$OH and the longest for $C_{16}$OH, see Eq. (15). The elastic modulus $G_0$ is also the greatest for $C_{12}$OH and the smallest for $C_{16}$OH. Possible explanation could be that in the presence of $C_{16}$OH the wormlike micelles are thicker (than those with $C_{12}$OH) and less flexible, which leads to greater persistence length, $L_p$, and smaller $G_0$; see Eq. (14).

It is also interesting that the values of $\eta_0$ and $\tau_R$ are greater for $C_{14}$OH despite the indications for longest micelles for $C_{12}$OH (see above). Insofar as $\xi \propto L_p^{2/5} L_e^{3/5}$, it seems that $\eta_0$ and $\tau_R$ are affected by changes in $L_p$ and $L_e$; see Eqs. (18) and (19).

**Table 2.** Rheological parameters for the systems of standard rheological behavior in Fig. 4d and 7b (salt curves): 12 wt% $C_{16,18}$SME + CAPB ($w = 0.2$) + 0, 10 and 20 mM added $C_{12}$OH at various total NaCl concentrations.

| NaCl (mM) | $\eta_0$ (Pa·s)* | $G_0$ (Pa) | $\tau_R$ (s) | $G_0\tau_R$ (Pa·s) | $\xi$ (nm) | $\bar{\zeta}$ | $\zeta$ | $\tau_{br}$ (s) | $\tau_{rep}$ (s) |
|---|---|---|---|---|---|---|---|---|---|
| 12 wt% $C_{16,18}$SME + CAPB + NaCl | | | | | | | | | |
| 345 | 59 | 147 | 0.402 | 59.1 | 30.5 | 2.38 | 1.132 | 0.957 | 0.85 |
| 1045 | 72 | 222 | 0.318 | 70.6 | 26.6 | 0.20 | 0.008 | 0.064 | 8.00 |
| 12 wt% $C_{16,18}$SME + CAPB + 10 mM $C_{12}$OH + NaCl | | | | | | | | | |
| 345 | 160 | 208 | 0.736 | 153 | 27.2 | 0.91 | 0.166 | 0.670 | 4.05 |
| 645 | 147 | 255 | 0.565 | 144 | 25.4 | 0.20 | 0.008 | 0.113 | 14.1 |
| 12 wt% $C_{16,18}$SME + CAPB + 20 mM $C_{12}$OH + NaCl | | | | | | | | | |
| 245 | 88 | 155 | 0.573 | 88.8 | 30.0 | 2.00 | 0.800 | 1.15 | 1.44 |
| 545 | 106 | 276 | 0.392 | 108 | 24.8 | 0.20 | 0.008 | 0.078 | 9.75 |

*The value of $\eta_0$ is taken from the plateau of the respective flow curve, like those in Fig. 2b.

Table 2 shows data for pairs of points from the salt curves in Fig. 4d, which are selected to be almost symmetrical (with close viscosities) on both sides of the peak. Indeed, to the left and right of the peak, respectively, wormlike and branched micelles are expected to form [25,44-48]. We are aware of the fact that the above procedure for calculation of $\xi$, $\bar{\zeta}$, $\zeta$, $\tau_{br}$ and $\tau_{rep}$ (see the beginning of Section 4.3) is rigorously applicable only to wormlike micelles. Nevertheless, it is interesting to see what kind of differences will be found for systems of (almost) the same total surfactant concentration and viscosity, but of (supposedly) different structures.

It should be noted that the solutions of 12 wt% $C_{16,18}$SME + CAPB at $w = 0.2$ contain 145 mM NaCl, which comes as admixture in the used surfactant samples; see Section 2.1. For this reason, the total NaCl concentrations in Fig. 4d and Table 2 are with 145 mM higher than the concentrations of added NaCl denoted in Fig. 8b, which shows the Cole-Cole plots for the systems in Table 2. For all these systems, the total NaCl concentration is $\geq 245$ mM and all of them exhibit standard rheological behavior (Fig. 8b).

As in Table 1, in Table 2 the values of $\eta_0$ and $G_0\tau_R$ are in excellent agreement, despite the fact that $\eta_0$ is determined from flow curves in steady shear regime (see e.g. Fig. 2b), whereas $G_0$ and $\tau_R$ are independently determined from the crossover points of the $G'(\omega)$ and $G''(\omega)$ dependencies obtained in oscillatory regime (see e.g. Figs. 6a and b). This agreement is present in spite of the circumstance that the systems on the two sides of the peak of the salt curve could contain micelles of different structure (wormlike and branched). As already discussed, this fact implies that (at not too high $\omega$) the stress relaxation for both kinds of structures represents a single exponential relaxation.

In Table 2, we see also that in each pair of NaCl concentrations the values of $G_0$ are systematically smaller (whereas those of $\tau_R$ – larger) for the point to the left of the peak of the salt curve as compared with the point to the right of the peak. We could hypothesize that the greater $G_0$ value to the right of the peak indicates the formation of branched micelles of higher aggregation number, resembling microgel particles; see e.g. Ref. [45].

It should be also noted that for the points to the right of the peak (related, supposedly, to branched micelles), the values of $\bar{\zeta}$ (and the deviations from Maxwellian behavior) are systematically smaller than those for the points to the left (related to entangled wormlike micelles); see Table 2 and Fig. 8b.

Moreover, for the points to the left of the salt-curve peak in Table 2 the values of $\eta_0$ are, in general, greater than those in Table 1 at the same total surfactant concentration. Despite some differences in the surfactant composition, we could attribute this difference mostly to the presence of added NaCl in the systems of Table 2. In general the addition of salt leads to greater $\bar{L}$ (longer micelles) and higher $\eta_0$; see Eq. (19). Having in mind this fact, it is interesting to note that the values of the breakage time, $\tau_{br}$, for the left-branch points in Table 2 are systematically greater than those in Table 1. We could attribute this effect to smaller rate constant of breakage, $k_1$, in the presence of added salt; see Eq. (15). Indeed, the added salt suppresses the electrostatic repulsion between the surfactant headgroups at the micelle surface, which is charge-charge repulsion for the ionic surfactant and dipole-dipole repulsion for the zwitterionic one. Then, the hydrophobic attraction in the micelle core better packs the monomers in the micelle and opposes its breakage (leads to smaller $k_1$ and greater $\tau_{br}$; see Table 2). In contrast, in the presence of stronger electrostatic repulsion in the headgroup region (no added salt) the micelle becomes a stressed (and partially stretched) structure, which is closer to its breakage limit; correspondingly, $k_1$ is greater and $\tau_{br}$ smaller (Table 1).

*4.4. Augmented Maxwell model*

In the case of large deviations from the Maxwell model, characterized by the values of $\bar{\zeta}$, the application of the reptation-reaction model faces the following limitations:

First, if $\bar{\zeta} > 7$, then Eq. (11) is not applicable and one cannot determine $\zeta$ and $\tau_{rep}$. The published statistical theories do not give any prescriptions about how to determine $\zeta$ and $\tau_{rep}$ in this case. Still, if $\tau_R$ is determined from the crossover point and $\bar{\zeta}$ is determined by comparison with the published computed values of $\bar{\zeta}$ [23,24,48], one can determine $\tau_{br} = \tau_R \bar{\zeta}$.

Second, if $\bar{\zeta} > 10.2$ (the largest value of $\bar{\zeta}$, for which computed curves have been published; see Ref. [24]), i.e. if the experimental points lie above the curve with $\bar{\zeta} = 10.2$ as in Figs. 8e and f, then $\bar{\zeta}$ cannot be determined by interpolation. Moreover, at such high deviations from Maxwellian behavior, $\tau_R$ and $G_0$ cannot be determined from the crossover

point – see Figs. 8c and d, and the related text. In such a case, the standard procedure of data processing cannot be applied to obtain any of the parameters $\tau_R$, $G_0$, $\bar{\zeta}$, $\zeta$, $\tau_{br}$, and $\tau_{rep}$.

Note that large deviations from the Maxwell model ($\bar{\zeta} > 10.2$) are frequently observed in the experiment. In fact, about a half of the experimental data reported in the present article belongs to this case. For example, as illustrated in Figs. 6c, d and f, significant deviations from Maxwellian behavior are observed at the same total surfactant concentration (12 wt%), at which other systems exhibit standard rheological behavior (Figs. 6a, b and e). "Nonstandard" rheological behavior is often observed, so that for this case it is also necessary to develop methodology for quantitative analysis and interpretation of the obtained data.

Here, we propose an augmented version of the Maxwell model, which allow one to determine at least $\tau_R$ and $G_0$ in the case of large deviations ($\bar{\zeta} > 10.2$). Moreover, the augmented Maxwell model allows one to characterize the viscoelastic medium with the frequency dependences of the mean elasticity and viscosity, $\langle G \rangle$ and $\langle \eta \rangle$, which are averaged over one oscillatory period.

For this goal, let us consider the Cole-Cole plots in Figs. 8e and f. For these plots, the deviation of the experimental points from the Maxwellian semicircle is so large, that it exceeds the deviation for the greater computed $\bar{\zeta}$ value published in the literature, viz. $\bar{\zeta} = 10.2$ in Ref. [24]. Nevertheless, as seen in the cited figures in the region of low $\omega$ (small $G'$) the experimental points lie on circle. By fit of these points with Eq. (9), one can determine the elastic modulus $G_0$, which characterizes the Maxwellian behavior of the system in this restricted low-frequency region. Eq. (9) contains only one adjustable parameter, $G_0$, which makes its determination from the data fit more reliable.

The augmented Maxwell model, which is a kind of viscoelastic thixotropic model, was first introduced and used to process experimental data for viscoelastic protein adsorption layers [86-88]. Here, it is adapted to the case of micellar systems. This model allows analysis of the $G'(\omega)$ and $G''(\omega)$ curves in Fig. 6 in the whole range of their variation. The basis of this model is again Eq. (6) where it is assumed [86]:

$$G = G(\dot{\gamma}), \quad \eta = \eta(\dot{\gamma}), \quad v_{ch}(\dot{\gamma}) \equiv \frac{G(\dot{\gamma})}{\eta(\dot{\gamma})} \tag{20}$$

The substitution of Eqs. (2) and (3) into Eq. (6) leads to:

$$(v_{ch}G'' + G')\cos(\omega t) + (v_{ch}G' - G'')\sin(\omega t) = G\omega\cos(\omega t) \tag{21}$$

In view of Eq. (20), the multiplication of Eq. (21) by $\sin(\omega t)$ and $\cos(\omega t)$, with a subsequent integration, yields [86]:

$$\frac{G''}{G'}\omega = \frac{2}{\pi}\int_0^\pi v_{ch}(\dot{\gamma}(x))\sin^2 x\, dx \tag{22}$$

$$G' + \frac{2G''}{\pi\omega}\int_0^\pi v_{ch}(\dot{\gamma}(x))\cos^2 x\, dx = \frac{2}{\pi}\int_0^\pi G(\dot{\gamma}(x))\cos^2 x\, dx \tag{23}$$

where $\dot{\gamma}(x) = \gamma_a \omega \cos x$, and $x \equiv \omega t$ is an integration variable. At given $G'(\omega)$ and $G''(\omega)$, Eqs. (22) and (23) form a system of two integral equations for determining the functions $G(\dot{\gamma})$ and $v_{ch}(\dot{\gamma})$. Following Refs. [86-88], we can give a much simpler description of the viscoelastic behavior of the system in terms of the mean values $\langle G \rangle$ and $\langle v_{ch} \rangle$, averaged over one oscillatory period. The quantities $\langle G \rangle$ and $\langle v_{ch} \rangle$ are defined as follows [86]:

$$\langle G \rangle \equiv \frac{2}{\pi}\int_0^\pi G(\dot{\gamma}(x))\cos^2 x\, dx \tag{24}$$

$$\langle v_{ch} \rangle \equiv \frac{2}{\pi}\int_0^\pi v_{ch}(\dot{\gamma}(x))\sin^2 x\, dx = \frac{G''}{G'}\omega \tag{25}$$

In the limit of the standard Maxwell model, we have $G \equiv$ const. and $v_{ch} \equiv$ const., and then Eqs. (24) and (25) reduce to $\langle G \rangle = G$ and $\langle v_{ch} \rangle = v_{ch} = G''\omega/G'$. The latter relation is consistent with Eq. (8), insofar as in the limiting case we have $v_{ch} = G_0/\eta_0$ (Maxwell model).

Following Ref. [86], we seek $v_{ch}(\dot{\gamma})$ in the form:

$$v_{ch}(\dot{\gamma}) = \frac{1}{\tau_R} + Q|\dot{\gamma}|^m \tag{26}$$

where $Q$ and $m$ are constants, and $1/\tau_R$ is the liming value of $v_{ch}(\dot{\gamma})$ at low shear rates. Substituting Eq. (26) in Eq. (25) and solving the integral, we obtain:

$$\langle v_{ch}\rangle = \frac{G''\omega}{G'} = \frac{1}{\tau_R} + \frac{1}{\tau_F}\left(\frac{\omega}{\omega_0}\right)^m \tag{27}$$

where, by definition $\omega_0 \equiv 1\ \text{s}^{-1}$ is the SI frequency unit and the quantity

$$\frac{1}{\tau_F} \equiv \frac{Q}{\sqrt{\pi}}\frac{\Gamma(m/2+0.5)}{\Gamma(m/2+2)}\gamma_a^m \omega_0^m \tag{28}$$

scales the frequency dependent part of $\langle v_{ch}\rangle$; $\Gamma(x)$ is the gamma function. Being a corollary from the empirical Eq. (26), Eq. (27) is in good agreement with the experiment; see Section 4.5.

Furthermore, in view of Eq. (24), the substitution of Eq. (26) in the left-hand side of Eq. (23), after solving the integral, leads to the expression:

$$\langle G\rangle = \frac{G'^2 + (m+1)G''^2}{G'} - \frac{mG''}{\tau_R \omega} \tag{29}$$

Having in mind the relation $\eta = G/v_{ch}$, we define $\langle \eta\rangle \equiv \langle G\rangle/\langle v_{ch}\rangle$. This definition, along with Eq. (29) and $\langle v_{ch}\rangle = G''\omega/G'$ yields:

$$\langle \eta\rangle = \frac{G'^2 + (m+1)G''^2}{G''\omega} - \frac{mG'}{\tau_R \omega^2} \tag{30}$$

For $m = 0$, Eqs. (29) and (30) reduce to Eq. (8).

If the system obeys the conventional Maxwell model for $\omega\to 0$, then in this limit $G' \approx G_0(\tau_R\omega)^2$ and $G'' \approx G_0\tau_R\omega$; see Eq. (4). In the same limit, Eqs. (29) and (30) yield:

$$\lim_{\omega\to 0}\langle G\rangle = G_0,\quad \lim_{\omega\to 0}\langle \eta\rangle = G_0\tau_R = \eta_0 \tag{31}$$

*4.5. Systems with nonstandard rheological behavior: numerical results*

The procedure for data processing is as follows:

(1) The data are plotted as $G''\omega/G'$ vs. $\omega$ and fitted with Eq. (27). The parameters $\tau_R$, $\tau_F$ and $m$ are determined from the fit. There are two alternative procedures for determining $G_0$, as follows:

(2a) $G_0$ is determined from the Cole-Cole plot by fit with Eq. (9) of the data in the region at low $\omega$ where the data comply with circle (see Figs. 8e and f). This procedure works well, unless the aforementioned region is too narrow.

(2b) $G_0$ is determined from the plot of $\langle G \rangle$ vs. $\omega$ as the limiting value for $\omega \to 0$; see Eq. (31) and Fig. 9c. This procedure could also work well, if the rheometer produces reliable data at the lower $\omega$ values.

One could determine $G_0$ by both procedures and compare the obtained results.

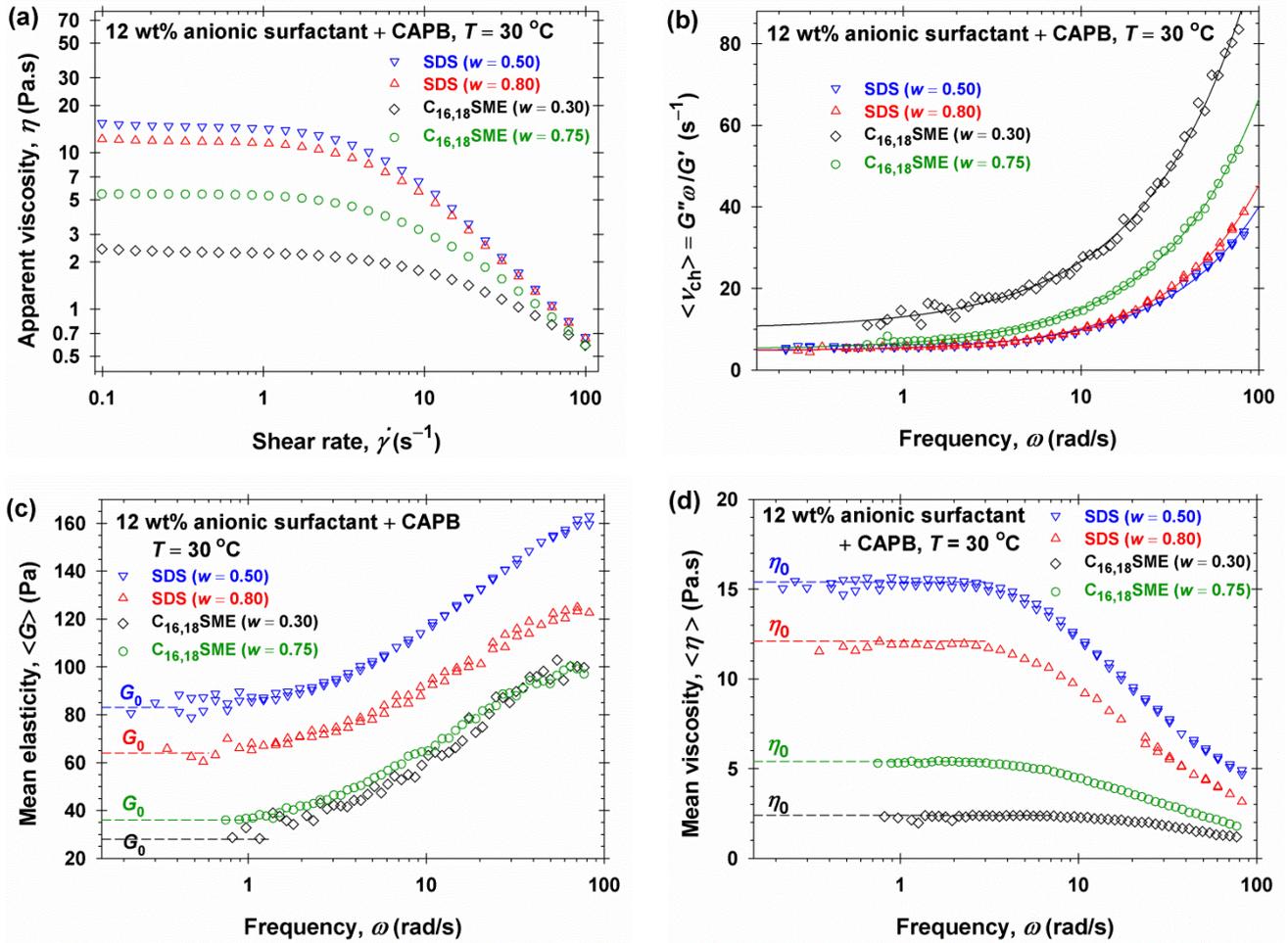

**Fig. 9.** Data processing for systems with nonstandard rheological behavior: (a) Flow curves in steady shear regime: plots of the apparent viscosity, $\eta$, vs. the shear rate, $\dot{\gamma}$. (b) Mean characteristic frequency $\langle v_{ch} \rangle$ vs. the frequency of oscillations, $\omega$; the solid lines are fits with Eq. (27). (c) Mean elasticity $\langle G \rangle$ vs. $\omega$ and (d) mean viscosity $\langle \eta \rangle$ vs. $\omega$ calculated from the experimental $G'$ and $G''$ using Eqs. (29) and (30).

Fig. 9a shows the flow curves obtained in steady shear regime for the systems with nonstandard rheological behavior in Figs. 6c and d. The flow curves in this figure are regular, like those in Figs. 2a and b.

Fig. 9b shows plots of the data for $G'(\omega)$ and $G''(\omega)$ in Figs. 6c and d as $\langle \nu_{ch} \rangle = G''\omega/G'$ vs. $\omega$. The obtained experimental curves are fitted with Eq. (27); the theoretical curves are in excellent agreement with the experimental dependencies. At that, $\tau_R$, $\tau_F$ and $m$ have been determined as adjustable parameters from the best fits. The obtained values are given in Table 3 together with $\eta_0$ determined from the plateaus of the flow curves at low $\dot{\gamma}$ in Fig. 9a, whereas $G_0$ is determined as the limiting value of $\langle G \rangle$ for $\omega \to 0$ (Fig. 9c); the correlation length $\xi$ is estimated from Eq. (13).

**Table 3.** Rheological parameters for the systems of nonstandard rheological behavior in Fig. 9: 12 wt% total surfactant concentration at different weight fractions of CAPB, $w$.

| System | $\eta_0$ (Pa·s)* | $G_0$ (Pa) | $\tau_R$ (s) | $G_0 \tau_R$ (Pa·s) | $\xi$ (nm) | $\tau_F$ (s) | $m$ |
|---|---|---|---|---|---|---|---|
| SDS + CAPB, $w = 0.50$ | 15.5 | 83 | 0.209 | 17.3 | 36.8 | 1.39 | 0.845 |
| SDS + CAPB, $w = 0.80$ | 12.1 | 64 | 0.214 | 13.7 | 40.1 | 1.34 | 0.868 |
| $C_{16,18}$SME + CAPB, $w = 0.30$ | 2.3 | 28 | 0.099 | 2.8 | 53.0 | 0.339 | 0.748 |
| $C_{16,18}$SME + CAPB, $w = 0.75$ | 5.3 | 36 | 0.198 | 7.1 | 48.8 | 0.620 | 0.792 |

*The value of $\eta_0$ is taken from the plateau of the respective flow curve in Fig. 9a.

For the data in Table 3, the relation $\eta_0 \approx G_0 \tau_R$ is better fulfilled for the more viscous systems. The data indicate that the characteristic time $\tau_F$, which scales the frequency term in Eq. (27), increases almost linearly with $G_0 \tau_R$ (with viscosity). For greater $\tau_F$, the contribution of the frequency term in Eq. (27) is smaller and the rheological behavior of the system is closer to Maxwellian – the curves in Fig. 9b closer approach a horizontal line of ordinate $\langle \nu_{ch} \rangle = 1/\tau_R$. The power $m$ varies in a relatively narrow interval, between 0.75 and 0.87.

The increase of $\nu_{ch}$ with the rise of $\omega$ (Fig. 9b) means that with the rise of angular frequency the system relaxes faster. The plots of the average elasticity and viscosity, $\langle G \rangle$ and $\langle \nu_{ch} \rangle$ vs. $\omega$ in Figs. 9c and d indicate that this behavior is accompanied by increase of elasticity and decrease of viscosity with the rise of $\omega$. At $\omega \to 0$, the values of $\langle G \rangle$ and $\langle \nu_{ch} \rangle$ approach $G_0$ and $\eta_0$, in agreement with Eq. (31). Note also that the shapes of the frequency dependences of elasticity $\langle G \rangle$ and storage modulus $G'$ are rather different; in particular, $G' \to 0$ whereas $\langle G \rangle \to G_0$ for $\omega \to 0$. In contrast, the shape of the dependence $\langle \eta \rangle$ vs. $\omega$ is similar to the dependence of apparent viscosity, $\eta$, on the strain rate, $\dot{\gamma}$; compare Figs. 9a and 8d. In particular, $\langle \eta \rangle \approx \eta$ in the plateau region at $\omega \to 0$. However, at higher $\omega$ values these curves differ, which is related to the different kinetic regimes: oscillations vs. steady shear.

From the viewpoint of data processing, the fit of the data for $G''\omega/G'$ vs. $\omega$ with Eq. (27) (as in Fig. 9b) provides a convenient way for determining $\tau_R$. The dependence of $\tau \equiv 1/\langle \nu_{ch} \rangle$ on $\omega$ (Fig. 9b) could be considered as an estimate for the mean relaxation time of a system with complex relaxation processes, which could involve several different characteristic relaxation times. Finally, the plots in Figs. 9c and d characterize the considered systems with frequency dependences of the elasticity and viscosity, $\langle G \rangle$ and $\langle \eta \rangle$, averaged over one oscillatory period, which should obey Eq. (31).

*4.6. Comparison of systems with standard and nonstandard behavior*

The experimental curves in Figs. 6e and f for 12 wt% $C_{16,18}$SME + CAPB with added 80 and 20 mM CMEA exhibit, respectively, standard and nonstandard rheological behavior. For the same systems, Fig. 10a shows the flow curves obtained in steady shear regime, which have regular shape. From the plateaus at low $\dot{\gamma}$, the values of $\eta_0$ in Tables 4 and 5 are determined.

Fig. 10b shows the Cole-Cole plots for the systems with standard rheological behavior (with 80 mM CMEA). For these systems, the values of $G_0$, $\tau_R$, $\xi$, $\bar{\zeta}$, $\zeta$, $\tau_{br}$ and $\tau_{rep}$, are determined as described in Section 4.3; see Table 4. For the system with $w = 0.2$, the greatest $\bar{\zeta} = 9$ was obtained. For this system $\zeta$ and $\tau_{rep}$ were not determined because Eq. (11) could be inapplicable at $\bar{\zeta} = 9$. The agreement of the independently determined $G_0\tau_R$ and $\eta_0$ is excellent (Table 4). The greater value of $\tau_{rep}$ for the system with $w = 0.80$ (located at the right end of the high "central plateau" in Fig. 4b) indicates the formation of long entangled wormlike micelles, which is evidenced also by the high values of $\eta_0$ and $G_0$ for this system. At $w = 0.85$ (to the right of the plateau), the viscosity drops by one order of magnitude and the shorter reptation time ($\tau_{rep} = 0.169$) evidences significant shortening of the micelles. The shorter $\tau_{br}$ for this system could indicate that these micelles are also more labile, i.e. the breakage constant $k_1$ in Eq. (15) is greater.

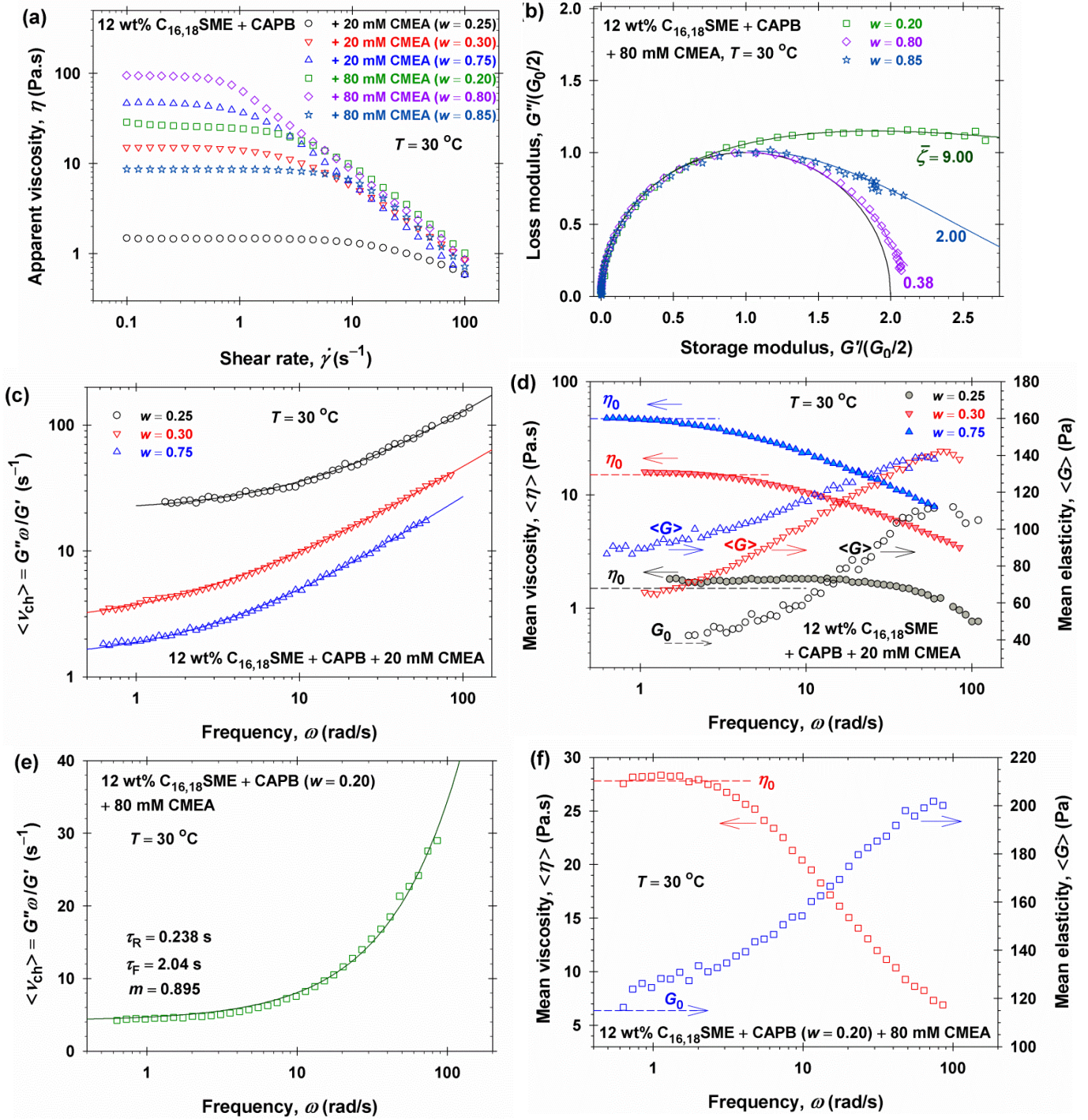

Fig. 10. Data processing for 12 wt% $C_{16,18}$SME + CAPB with 20 and 80 mM added CMEA, at various weight fractions of CAPB, $w$. (a) Flow curves in steady shear regime: plots of $\eta$ vs. $\dot{\gamma}$. (b) For 80 mM CMEA at $w$ = 0.20, 0.80 and 0.85, the Cole-Cole plots show that the data comply with the Maxwell model extended with Cates theory at greater $G'$ values. (c) Plot of $\langle v_{ch} \rangle$, vs. $\omega$ for 20 mM CMEA at $w$ = 0.25, 0.30 and 0.75; the solid lines are fits with Eq. (27). (d) Plots of the mean viscosity $\langle \eta \rangle$ and elasticity $\langle G \rangle$ vs. $\omega$ calculated from the experimental $G'$ and $G''$ using Eqs. (29) and (30). (e) Fit of the data for $\langle v_{ch} \rangle$ vs. $\omega$ for the systems with $\bar{\zeta} = 9.00$ by means of the augmented Maxwell model, Eq. (27), and (f) plots of $\langle \eta \rangle$ and $\langle G \rangle$ vs. $\omega$ for the same system.

**Table 4.** Rheological parameters for the systems of standard rheological behavior in Fig. 10b: 12 wt% $C_{16,18}$SME + CAPB + 80 mM CMEA at different weight fractions of CAPB, $w$.

| CAPB | $\eta_0$ (Pa·s) | $G_0$ (Pa) | $\tau_R$ (s) | $G_0\tau_R$ (Pa·s) | $\xi$ (nm) | $\bar{\zeta}$ | $\zeta$ | $\tau_{br}$ (s) | $\tau_{rep}$ (s) |
|---|---|---|---|---|---|---|---|---|---|
| $w = 0.20$ | 27.8 | 115 | 0.238 | 27.4 | 33.1 | 9.00 | – | 2.14 | – |
| $w = 0.80$ | 92.3 | 232 | 0.394 | 91.4 | 26.2 | 0.38 | 0.029 | 0.150 | 5.20 |
| $w = 0.85$ | 8.63 | 128 | 0.0676 | 8.65 | 32.0 | 2.00 | 0.800 | 0.135 | 0.169 |

*The value of $\eta_0$ is taken from the plateau of the respective flow curve in Fig. 10a.

**Table 5.** Rheological parameters for the systems of nonstandard rheological behavior in Fig. 10c: 12 wt% $C_{16,18}$SME + CAPB + 20 mM CMEA at different weight fractions of CAPB, $w$.

| CAPB | $\eta_0$ (Pa·s) | $G_0$ (Pa) | $\tau_R$ (s) | $G_0\tau_R$ (Pa·s) | $\xi$ (nm) | $\tau_F$ (s) | $m$ |
|---|---|---|---|---|---|---|---|
| $w = 0.25$ | 1.50 | 38 | 0.0481 | 1.83 | 47.8 | 0.489 | 0.860 |
| $w = 0.30$ | 15.1 | 40 | 0.394 | 15.8 | 47.0 | 0.828 | 0.783 |
| $w = 0.75$ | 47.0 | 65 | 0.727 | 47.3 | 40.0 | 1.97 | 0.852 |

*The value of $\eta_0$ is taken from the plateau of the respective flow curve in Fig. 10a.

For the systems with 20 mM CMEA and nonstandard rheological behavior, Fig. 10c shows plots of the data and fits with Eq. (27). The values of $\tau_R$, $\tau_F$ and $m$ determined from the fits are given in Table 5, which contains also the values of $G_0$ obtained from Cole-Cole plots like those in Fig. 8e and f. Once again, the independently determined $\eta_0$ and $G_0\tau_R$ are in very good agreement. Among these systems, the viscosity and elasticity, $\eta_0$ and $G_0$, are the highest (and the formed micelles should be the longest and the most flexible) for the system with $w = 0.75$, which is situated slightly to the right of the central plateau in Fig. 4b. In Fig. 10c, for this system, $\langle v_{ch} \rangle$ is the lowest, which means that the average relaxation time $\tau \equiv 1/\langle v_{ch} \rangle$ is the longest, as expected for the system of the highest viscosity. Moreover, the values of $\tau_F$ in Table 5 increase almost linearly with $\eta_0$. The values of $m$ are in the range between 0.78 and 0.86. With the determined $\tau_R$ and $m$, the mean viscosity and elasticity, $\langle \eta \rangle$ and $\langle G \rangle$, are calculated from Eqs. (29) and (30) and plotted vs. $\omega$ in Fig. 10d. At $\omega \to 0$, the averaged viscosity approaches the respective Maxwellian value: $\langle \eta \rangle \to \eta_0$, whereas fulfillment of the relation $\langle G \rangle \to G_0$ is observed only for the lowest experimental curve. (For the other two curves, data at lower $\omega$ are needed, but they cannot be obtained with the used rheometer.)

The augmented Maxwell model can be applied also to systems with standard rheological behavior. As an example, let us consider the system with 12 wt% $C_{16,18}$SME + CAPB at $w = 0.2$ with 80 mM added CMEA, for which $\bar{\zeta} = 9$ (Table 4). In Fig. 10e, the data for $G'(\omega)$ and $G''(\omega)$ for this system (see Fig. 6e) are plotted as $\langle v_{ch} \rangle = G''\omega/G'$ vs. $\omega$, and excellent agreement with Eq. (27) is obtained. The flat part of the curve (at lower $\omega$ values), where $\langle v_{ch} \rangle \approx 1/\tau_R$ = const., corresponds to the region of Maxwellian behavior of the micellar system, with single relaxation time $\tau_R$. The increasing part of the curve in Fig. 10e at higher $\omega$ corresponds to rheological response that is characterized with either milti-exponential or non-exponential stress relaxation. Finally, Fig. 10d shows the calculated (from Eqs. (29) and (30)) dependences of $\langle \eta \rangle$ and $\langle G \rangle$, whose limiting values for $\omega \to 0$ are in agreement with Eq. (31).

The augmented Maxwell model is appropriate for processing experimental data for $G'(\omega)$ and $G''(\omega)$ in the cases of not too small deviations from the conventional Maxwellian behavior, i.e. for not too small values of $\bar{\zeta}$. In the opposite case of small $\bar{\zeta}$, the plot of $\langle v_{ch} \rangle = G''\omega/G'$ vs. $\omega$ is close to a horizontal line ($\langle v_{ch} \rangle \approx 1/\tau_R$ = const.), from which it is impossible to accurately determine $\tau_F$ and $m$ by fit with Eq. (27). In this case, the data processing procedure for systems with standard rheological behavior (Section 4.3) works very well. In the general case, when analysis of raw data for $G'(\omega)$ and $G''(\omega)$ is carried out, the plot of $\langle v_{ch} \rangle = G''\omega/G'$ vs. $\omega$ allows one to identify the frequency domains with Maxwellian and non-Maxwellian behavior (with constant and varying $\langle v_{ch} \rangle$), and to determine $\tau_R$.

## 5. Conclusions

The mixing of anionic and zwitterionic surfactants in aqueous solutions is known to promote the synergistic growth of giant micelles, which is detected as a significant rise of viscosity [1,10,29,31,39,42,50-54]. In the present paper we investigated this phenomenon for mixed solutions of sulfonated methyl esters, SME, and CAPB, potential ingredients for personal-care formulations. The effect of SME is compared with those of a standard anionic surfactant, SDS. Moreover, the effect of additives, such as fatty alcohols and CMEA, which are expected to boost the micelle growth and act as thickening agents, is studied. The impact of added salt, which could either increase or decrease the viscosity (depending on its concentration), is also investigated.

A complete systematic study on the role of all these components considerably exceeds the scope of the presents study. Here, we report the most significant observed effects with an emphasis on the interpretation of the obtained rheological data and understanding at molecular level the processes and phenomena that give rise to the observed macroscopic rheological behavior.

The most important new experimental finding is that the mixing of SME and CAPB produces a significant synergistic effect manifested by the rise of viscosity. Without or with added salt, the effect of SME is stronger than that of SDS (Fig. 4a). The addition of thickening agents, such as fatty alcohols, CMEA and cationic polymer, leads to broadening of the synergistic peak in viscosity without significantly affecting its height (Fig. 4b).

The addition of NaCl leads to a typical salt curve with high maximum (Fig. 4d). Surprisingly, the addition of fatty alcohol (dodecanol) considerably decreases the height of the salt-curve peak and shifts it to the left (to lower salt concentrations). Thus, it turns out that at lower salt concentrations the alcohol acts as a thickener, whereas at higher salt concentrations – as a thinning agent.

Among the studied alcohols, the strongest thickening effect is produced by dodecanol and tetradecanol (Figs. 5a and b). The viscosity increases with the alcohol concentration up to a certain limit, at which the solutions become turbid and their viscosity decreases. The observed precipitation is related to the transformation of the giant micelles into drops or crystallites. The addition of the nonionic surfactant CMEA does not cause precipitation, but promotes the micelle growth and viscosity rise (Fig. 5c).

The above conclusions are based on viscosity measurements in steady shear regime. In addition, experiments in oscillatory regime were carried out, which yielded the frequency dependences of the storage and loss moduli, $G'(\omega)$ and $G''(\omega)$. Depending on the shape of the latter dependences, one can distinguish systems with standard and nonstandard rheological behavior (Fig. 6). The systems with standard behavior obey the Maxwell viscoelastic model (at least) up to the crossover point ($G' = G''$), which allows determination of the Maxwell elastic modulus, $G_0$, and characteristic relaxation time, $\tau_R$. Moreover, the Cates reptation-reaction model [19,23,25] can be applied to characterize such systems with the micelle breakage and reptation times, $\tau_{br}$ and $\tau_{rep}$ (see Section 4.2 and Tables 1, 2 and 4).

The systems with nonstandard rheological behavior obey the Maxwell model only in a restricted frequency domain below the crossover frequency $\omega_c$ (Fig. 8e and f). It is demonstrated that the rheological data for such systems could be interpreted in the framework of an augmented version of the Maxwell model, which adopts dependence of elasticity and viscosity on the shear rate: $G(\dot{\gamma})$ and $\eta(\dot{\gamma})$ (see Section 4.4). It is convenient to determine the elastic modulus $G_0$ from the Cole-Cole plot (Figs. 8e and f) and the relaxation time $\tau_R$ – from the plot of the mean characteristic frequency $\langle \nu_{ch} \rangle = G''\omega/G'$ vs. $\omega$ (Figs. 9b; 10c and e). The latter plot allows one to identify the frequency domains with Maxwellian and non-Maxwellian behavior, and to determine also the values of two other parameters of the augmented model – $\tau_F$ and $m$. Furthermore, one can characterize the rheological properties of the system with the frequency dependences of the mean elasticity and viscosity, $\langle G \rangle$ and $\langle \eta \rangle$; see Figs. 9c and d, and 10d and f. The correctness of the proposed theoretical interpretation (for systems of both standard and nonstandard rheological behavior) is supported by the good agreement of the values of $\eta_0$ and $G_0\tau_R$, which have been determined from independent sets of data obtained in the steady-shear and oscillatory regimes; see Tables 1–4.

Standard rheological behavior is observed with systems of higher viscosity (and, supposedly, longer micelles) in the presence of added alcohol and/or salt, or CMEA at higher concentration; see Tables 1, 2 and 4. Nonstandard rheological behavior is typically observed for systems of not too high viscosity (below $\eta_0$ = 50 Pa·s) in the absence (or at lower concentrations) of thickeners. However, in terms of viscosity there is no sharp border between systems of standard and nonstandard behavior (compare, e.g., Tables 4 and 5).

Possible explanation of the viscosity peak observed upon the variation of anionic/zwitterionic surfactant composition (Figs. 4a and b) could be the growth and diminishing of the length of wormlike micelles due to synergistic interactions of the two types of surfactant headgroups [10]. The increase of viscosity upon the addition of fatty alcohols and CMEA (Fig. 5) could be interpreted as a result of additional increase of the wormlike micelle length. Instead, the peak in the salt curves (Fig. 4d) could be interpreted as a transition from wormlike to branched micelles; see e.g. Refs. [44,45,52]. In future studies, these micellar structures and shape transformations have to be confirmed by cryogenic transmission electron microscopy (cryo-TEM) [89].

This paper could be useful for experimentalists and industrial researchers who are optimizing their experimental systems and formulations, and for theoreticians who are working on testing and extension of their models. Future development of the present study could include completion of the systematic investigation of rheological effects of various additives at various concentrations on the rheology of mixed SME + CAPB solutions and other similar systems, as well as further extension and application of the mean-field models for physical interpretation of the properties of systems with both standard and nonstandard rheological behavior.

**Acknowledgments**

All authors gratefully acknowledge the support from KLK OLEO. VY, KD and PK acknowledge the support from the Operational Programme "Science and Education for Smart Growth", Bulgaria, grant number BG05M2OP001-1.002-0012. GR acknowledges the financial support received from the program "Young scientists and postdoctoral candidates" of the Bulgarian Ministry of Education and Science, MCD № 577/17.08.2018.